\documentclass[12pt,a4paper]{article}

\usepackage{amsmath,amsthm,amssymb}
\usepackage{graphics,graphicx}
\usepackage{epsfig}
\makeatletter
\@addtoreset{equation}{section}
\makeatother

\begin{document}
\begin{flushright}
QMUL-PH-05-11\\
\end{flushright}
\begin{center}
\Large \textbf{Fuzzy Sphere Dynamics and Non-Abelian DBI in Curved Backgrounds.}\\
\vspace{1.5cm}
\normalsize \textbf{Steven Thomas\footnote{s.thomas@qmul.ac.uk} and
John Ward\footnote{j.ward@qmul.ac.uk}}\\
\vspace{1.5cm}
\normalsize \emph{ Department of Physics \\
Queen Mary, University of London \\
Mile End Road, London \\ E1 4NS, U.K \\}
\vspace{1.5cm}
\textbf{Abstract}
\end{center}
We consider the non-Abelian action for the dynamics of $N Dp'$-branes in
the background of $M Dp$-branes, which parameterises a fuzzy sphere using the $SU(2)$
algebra. We find that the curved background leads to collapsing solutions for the fuzzy sphere
except when we have $D0$ branes in the $D6$ background, which is a realisation of the gravitational Myers effect.
Furthermore we find the equations of motion in the Abelian and non-Abelian theories are identical in the
large $N$ limit.
By picking a specific ansatz we find that we can
incorporate angular momentum into the action, although this imposes restriction upon the dimensionality
of the background solutions. We also consider the case of non-Abelian non-BPS branes, and examine the
resultant dynamics using world-volume symmetry transformations. We find that the fuzzy sphere always
collapses but the solutions are sensitive to the combination of the two conserved charges and we
can find expanding solutions with turning points.
We go on to consider the coincident $NS$5-brane background, and again construct the non-Abelian
theory for both BPS and non-BPS branes. In the latter case we must use symmetry arguments
to find additional conserved charges on the world-volumes to solve the equations of motion.
We find that in the Non-BPS case there is a turning solution for specific regions of the tachyon and radion fields.
Finally we investigate the more general dynamics of fuzzy $\mathbb{S}^{2k}$ in the $Dp$-brane background, and find
collapsing solutions in all cases.\\

\newpage
\section{Introduction}
There has been much recent work on time dependence in gravitational backgrounds 
\cite{branonium, time_dependence}.
The basic idea has been to introduce a probe BPS $Dp$ brane into the non-trivial geometry
of a large number of background branes and study its corresponding dynamics. This has also
been extended to include non-BPS branes and supertube probes. The introduction of a probe brane tends to break all the
supersymmetry associated with the background configurations, and therefore the probe will experience
a gravitational force due to the source branes. Of course, by selecting specific probes in backgrounds
we can preserve the superymmetries and there will be no net force. However we generally see that probes placed
in the non-trivial backgrounds are unstable, and share many similarities to the condensation of the open string
tachyon \cite{sen, non_bps_dynamics}. In particular, it can be seen that the energy momentum tensor localised on the probe brane has
vanishing pressure at late times which is similar to the fluid at the tachyonic vacuum \footnote{Recall that
the open string degrees of freedom at the tachyon vacuum vanish and only closed string modes remain. This is
due in part to the reduction of the metric to a Carollian form \cite{tachyon_stuff}.}.

It has also been suggested that the open string tachyon may have a geometrical interpretation in terms of 
one dimensional brane motion in a confined, bounded non-trivial background with $\mathbb{Z}_2$ symmetry 
\cite{geometrical_tachyons, kinks}.
Specifically we see that the radion field, parameterising
the distance of a probe brane from the source branes, becomes tachyonic when placed at the unstable point in the background. In addition, we
have seen that these geometrical tachyons exhibit similar kink solutions to those of the open string tachyon, and
we would also expect there to be stable vortex solutions. Although this has proven to be tantalising, it still
remains to be seen whether it is possible to determine the true relationship between the open string tachyon
and its geometrical cousin.

Most of this work, however, has focused on a solitary probe brane thus it seems logical that this program should be
extended to include multiple branes. As is well known, the presence of N coincident $Dp$-branes implies that
there is a unitary $U(N)$ gauge theory, due to the open string degrees of freedom \cite{myers}. This means that the 
effective DBI action is no longer applicable and we must resort to using the non-Abelian extension \cite{myers, tseytlin}.
One of the major differences between the two is that the scalar fields must now transform under a representation
of a gauge group. Therefore they no longer commute with one another, leading us to introduce the notion of non-commutative
coordinates and hence many of the ideas associated with non-commutative geometry. Although this approach has been useful, it
is known that the non-Abelian action agrees only up to terms of order $F^6$ \cite{myers} when compared to exact open string calculations.
Furthermore, there has been no satisfactory resolution to the problem of the finite $N$ expansion of the action.
Despite this, there has been an incredible amount of work done in this field with regards to intersecting
brane configurations leading to the construction of fuzzy funnels. One of the byproducts of this has been the
large-small dualities between funnel solutions and collapsing spheres sourced by $D0$-branes \cite{costis}. Again it seems only
logical to look at non-trivial backgrounds to see if these dualities still hold. It has also been suggested that the event horizon of
black holes should be described by fuzzy spheres. If this is the case, then our analysis would hopefully yield some solutions
with regard to the classical stability of such as system.

This paper will attempt to analyse the dynamics of several probe branes in curved backgrounds of coincident $D$-branes
and $NS$5-branes using the irreducible representation of $SU(2)$, which corresponds to a fuzzy sphere geometry.
We will only consider flat static branes all localised at the same point in the bulk space-time. More complicated
backgrounds such as the ring configuration will not be analysed \cite{sfetsos, israel}, although should be tackled at some point in the future.
One of the most important things to note is that there are exact conformal field theories associated with coincident
background solutions, and so any results obtained here will correspond to operators in the CFT.
We begin by constructing the low energy action for coincident $Dp'$-branes in a $Dp$-brane background, and examine
the solutions.
\section{Background solution and brane action.}
We consider the standard type II supergravity background solution for $M$ coincident
$Dp$-branes. These source branes are all assumed to be parallel in the sense that
their world volumes are oriented in the same directions, and that they are static. This will ensure that our
solutions are as simple as possible.
The 10-dimensional bulk spacetime is assumed to be infinite in extent, and there are no gravitational moduli in the
problem. The solutions for the metric, dilaton and R-R field are given by \cite{branonium, stelle}
\begin{eqnarray}
ds^2&=&H^{-1/2}\eta_{\mu \nu} dx^{\mu} dx^{\nu} + H^{1/2} dx^m dx^n \nonumber \\
e^{\phi} &=& H^{(3-p)/4} \nonumber \\
C_{0 \ldots p} &=& 1-H^{-1},
\end{eqnarray}
where $\mu, \nu$ represent directions parallel to the background branes, whilst $m, n$ are
transverse directions.
The harmonic function $H$ satisfies the Laplace equation in the transverse Euclidean space. In general it can be
written as a multi-centred function of the transverse coordinates:
\begin{equation}
H= 1 + \sum_{i=1}^M \frac{\tilde{k}_p}{|\textbf x-\textbf x_{i}|^{7-p}}
\end{equation}
which for coincident $D$-branes reduces to
\begin{equation}
H = 1 + \frac{k}{r^{7-p}}.
\end{equation}
where, $r=\sqrt{x_m x^m}$ and
$k_p=(2 \sqrt{\pi})^{5-p} M \Gamma(\frac{7-p}{2}) g_s l_s^{7-p}$.
As usual $l_s$ is the string length and $g_s$ is the string coupling at infinity.

Into this background we wish to insert $N$ probe $Dp'$-branes where we must ensure that
$N<M$ and also that $p \ge p'$ in order to satisfy the supergravity constraints (note that we will neglect the case of
$p'=-1$ in IIB, which corresponds to the D-instanton).
Because there is more than a single probe brane we can no longer use the Abelian DBI action, as
the extra massless string modes enhance the gauge symmetry on the world-volume. In order
to proceed we must first introduce the non-Abelian action for the bosonic fields.
The first part is the enhanced Born-Infeld contribution,
\begin{equation}\label{eq:actiondef}
S_{BI} = - \tau_p \int d^{p'+1}\zeta STr e^{-\phi} \sqrt{-det({\mathcal P}[E_{ab}+E_{ai}(Q^{-1}-\delta)^{ij}\\
E_{jb}]+\lambda F_{ab})}\sqrt{det(Q^i \ _j)}.
\end{equation}
where we have the usual definitions
\begin{equation}
\lambda = 2 \pi l_s^2, \hspace{1cm} E_{\mu \nu} = G_{\mu \nu}+B_{\mu \nu}, \hspace{0.5cm} \rm{and}
\hspace{0.5cm} Q^i \ _j \equiv \delta^i \ _j + i \lambda[ \phi^i, \phi^k] E_{kj}.
\end{equation}
The second part of the action is the Chern-Simons part coupling the background R-R field to the probe branes world volumes.
\begin{equation}
S_{CS}=\mu_{p} \int STr(\mathcal P[e^{i\lambda i_{\phi} i_{\phi}}\sum C^{(n)}e^B]e^{\lambda F}).
\end{equation}
As usual $\mathcal P[\ldots]$ represents the pullback of the spacetime tensors to the brane worldsheet.
The action contains $\phi_i$ terms, where $i=p+1 \ldots 9$ run over the
transverse coordinates. In fact these are the transverse scalars in the action which
are actually $N \times N$ matrix representations of the $U(N)$
worldsheet symmetry. The $STr(\ldots)$ denotes the symmetrized trace operation, the prescription for which is to take a symmetrized
average over all the possible orderings of the $F_{ab}, D_a \phi^i, i[\phi^i, \phi^k]$,
and all the possible orderings of the individual scalars prior to taking the trace.\footnote{In \cite{bordalo}, 
two loop corrections to the DBI action 
resulting from the curved background were computed. These lead to modifications of the symmetrized trace prescription   
and it would be of interest to see if this results in modifications to our fuzzy solutions.}  
In the Chern-Simons action we see the $i_\phi$ denote the interior product
by $\phi^i$ regarded as a vector in the transverse space. For a general $p$-form, we
see the interior derivatives act as
\begin{equation}
i_\phi i_\phi C^{(p)} = \frac{1}{2} [\phi^i, \phi^j] C_{ji}^{(p)}.
\end{equation}
It is well known that a $Dp$-brane is electrically charged under the $(p+1)$ form RR potential, with
a charge $\mu_p$. Supersymmetry constraints impose the additional condition that $\mu_p= \pm \tau_p$.
The non-Abelian Chern-Simons action shows that a $Dp$-brane can couple to R-R charges of higher
dimensionality, and thus permits the possibility of a brane dielectric effect.
For example, if we expand the Chern-Simons action to leading order
with no gauge field or $B$ field, we have
\begin{equation}
S_{CS}=\mu_p \int Tr( \mathcal{P} [C^{(p+1)}+i \lambda i_\phi i_\phi C^{(p+3)} - \frac{\lambda^2}{2}
(i_\phi i_\phi)^2C^{(p+5)}]).
\end{equation}
In this note we are assuming that all the probe branes are parallel to the source branes, therefore we find that the leading order
contribution to the Chern-Simons coupling reduces to:
\begin{equation}
S_{CS} = \mu_p \int Tr (\mathcal{P}[C^{p+1}])
\end{equation}
which, upon insertion of the background solutions, becomes
\begin{equation}
S_{CS} = -q\int dt N H^{-1}
\end{equation}
up to an arbitrary constant, where $q=+1$ corresponds to a $D$-brane probe and
$q=-1$ corresponds to an antibrane. Now, in the Abelian case we know that there
is only a coupling if $p=p'$ or if $p=6, p'=0$. Since we are neglecting
higher order corrections to the Chern-Simons action, we effectively have the same situation and
so we must remember to include these couplings in our effective theory.

To simplify the analysis as much as possible we will only consider time dependent
solutions for the transverse scalars. This will ensure that no caustics form in the
action. We will also set $F_{ab}$ to zero, and allow the only fields to be excited on the branes
to be those which are not in the angular directions. This will also ensure that the $B$
field will drop out of the action.
To ensure that the action is dimensionally consistent, we must be aware that the
$x_i$ (i=$p+1 \dots 9$) coordinates transform as
\begin{equation}
x_i = \lambda \phi_i,
\end{equation}
and the physical distance between background branes and probe branes in the harmonic function becomes
\begin{equation}
r^2 = \frac{\lambda^2}{N}Tr( \phi^i \phi^j \delta_{ij}).
\end{equation}
Now that we have set the stage, we can use our $Dp$-brane solutions to determine the dynamics of a collapsing fuzzy
sphere in this background, which we assume can be regarded as a probe of the geometry. Therefore we are neglecting
any back reaction and $1/N$ corrections in what follows.
\section{Radial Collapse.}
In this section, we will consider the purely radial motion of the $N$ $Dp'$-branes
in the background of the $M$ $Dp$-branes, where we must ensure that $M>N$ for the
supergravity solutions to hold. To simplify the problem even further, we
set all the coordinates to zero with the exception of $x_7 \ldots x_9$.
For simplicity we will only examine the $p=p'$ case in detail, as there are difficulties associated with the solutions
when $p\ne p'$. This shouldn't be surprising as the same thing happens in the Abelian case, where we must look for
world-volume symmetry transformations in order to solve the equations of motion. We expect this to hold in the non-Abelian
case, which poses questions about the relationship between non-Abelian brane solutions and the space-time uncertainty
principle. Although we will not discuss the implications in this note, it would certainly be interesting for future
investigations.
\subsection{Dynamics in the $p=p'$ case.}
In this particular instance, the background solution allows us to write the action as follows;
\begin{displaymath}\label{eq:action2}
S=-\tau_{p'} \int d^{p'+1} \zeta STr\left( H^{-1} \sqrt{(1-H \lambda^2 \dot{\phi^i} \dot{\phi^j}\\
\delta_{ij})({1-\frac{1}{2} \lambda^2 H [\phi^i, \phi^j][\phi^j,\phi^i])}} \right)\\
\end{displaymath}
\begin{equation}
S_{CS}=-\tau_{p'} \int d^{p'+1} \zeta \frac{qN}{H},
\end{equation}
where we have made the approximation $Q^{ij} \sim \delta^{ij}$, and only expanded the second
square root term to leading order. Our approximation that the inverse matrix $Q^{ij}$ is  treated as
unity to leading order in lambda  is consistent as long as our solution
only probes distances greater than the string length. As the fuzzy sphere
radius starts approaching $l_s$ we anticipate that higher order terms in
$Q^{ij}$ (and in the square root of det(Q) ) would need to be kept for
consistency. This approximation has been used by other authors who have
investigated fuzzy spheres in the nonabelian DBI theory, see for example the second  
paper in \cite{myers}.

In order to simplify the expression to something more useful we need to expand the commutator terms.
The simplest ansatz possible is to make the transverse scalars all commuting,
however it has been shown that the system will be unstable since
it will not be at its minimal energy. This can be easily be verified by expanding out the
last term in the action \cite{myers}.
Instead we opt for the more familiar $SU(2)$ ansatz which parameterises a
non-commutative object known as a fuzzy 2-sphere. The definition of which can be seen via
\begin{equation}\label{eq:fuzzyansatz}
\phi^i= R(t) T^i, \hspace{0.5cm} i=1, 2, 3
\end{equation}
where the $T^i$ are an $N \times N$ matrix representation of the generators of the $SU(2)$ algebra.
\begin{equation}
[T^i, T^j] = 2i \varepsilon_{ijk} T^k
\end{equation}

The remaining fields $\phi^i,  i= 4,5...$ are set to zero or more gemerally to constant matrices that commute 
with the $SU(2)$ generators.Let us make some comments concerning the generality and validity of this
'round'fuzzy sphere ansatz in (\ref{eq:fuzzyansatz}).  Our ansatz  sets  the  nonabelian transverse fields $\phi^i $
either to be $SU(2)$ valued fields (the fuzzy sphere ansatz) or to constant commuting
matrices. The latter are taken to commute with both the $SU(2)$ generators and
themselves. These latter fields have no potential because of they commute with everything, so the
assumption that they are constant is consistent with their equations of motion; they simply 
parameterise flat directions of the theory.   There is a  related issue of what is the
most general time dependent configuration..which is very interesting question.
For example one could imagine that there will be  non-spherical fluctuations
because there are a tidal effects in the direction of motion in the curved
backgrounds which should alter the geometry of  the fuzzy sphere...maybe leading to a fuzzy
'egg' . But these are deformations of  the spherical solution ..so we would
argue that in the first instance one should  study the latter first and then
investigate fluctations on about this solution.  There are other known fuzzy geometries with
different topology such as fuzzy cylinders which one could also investigate
in the context of curved backgrounds..but again this is outside the  remit of our
paper which focusses on spherical solutions. 

To check that our speherical ansatz is at least a consistent one we consider the equations of motion for
the nonabelian fields $\phi^i $ in a general curved background. Let us consider a background metric of the form
\begin{equation}\label{eq:metric}
ds^2 = -g_{00}dt^2 + g_{xx}dx^a dx^b \delta_{ab} + g_{zz} dz^i dz^j \delta_{ij}
\end{equation}
where $a, b$ run over the $q$ worldvolume directions and $i, j$ are transverse directions to the source. This background could obviously
be generated by a stack of coincident branes, or something more exotic. 
The non-Abelian action then take sthe form 
\begin{equation}\label{eq:nonabaction}
S =-\tau_p' \int d^{p'+1} \zeta STr \left(e^{-\phi}\sqrt{g_{xx}^p g_{00} (1-\lambda^2 g_{zz}g_{00}^{-1}\dot{\phi^i}\dot{\phi^j}\delta_{ij})} \sqrt{1-\frac{1}{2} \lambda^2 g_{zz}^2 [\phi^i, \phi^j][\phi^j,\phi^i] } \right)
\end{equation}
Note that restricting the metric components $g_{00}=g_{xx}=g_{zz}^{-1} = H^{1/2} $ the above action reproduces that in 
(\ref{eq:action2}) above. Now working to leading order in $\lambda$ the equations of motion for $\phi^i$ are
\begin{equation}\label{eq:eqmotion}
\frac{d}{dt} (e^{-2\phi}g_{xx}^{p/2}g_{00}^{-1/2}g_{zz} \dot{\phi^i} ) = g_{zz}^2 [\phi^i,[\phi^j,\phi^i]]
\end{equation}
Now consider the more general ansatz for $\phi^i $ 
\begin{equation}\label{eq:genansatz}
\phi^i= R(t) T^i + \beta(t) Y^i, \hspace{0.5cm} i=1, 2, 3
\end{equation} 
where the matrices $Y^i$ represent some non-spherical orthogonal directions to the $SU(2)$ generators $T^i $. 
Without loss of generality we can assume that $Tr(T^i Y^i) =0$. Using this property one can easily obtain equations of motion for $R(t) $ and 
$\beta(t)$ by substituting the above ansatz into (\ref{eq:eqmotion}). In the limit when we send $\beta(t)=0$ (ie our spherical fuzzy sphere ansatz)
the equation of motion for $\beta(t) $ becomes
\begin{equation}\label{eq:eqmotion1}
\frac{d}{dt} (e^{-2\phi}g_{xx}^{p/2}g_{00}^{-1/2}g_{zz} \dot{\beta} ) = \frac{1}{Tr(Y^iY^j)}g_{zz}^2Tr( [T^i,[T^j,T^i]] Y^j)
\end{equation}
Due to the orthogonality of $T^i$ and $Y^j$ the second trace factor in (\ref{eq:eqmotion1}) vanishes so 
$e^{-2\phi}g_{xx}^{p/2}g_{00}^{-1/2}g_{zz} \dot{\beta} $ is a constant. We can choose this constant to be zero and hence $\dot{\beta}$ also 
vanishes. It is therefore consistent to set $\beta(t)=0 $ at the outset as in our spherical fuzzy sphere ansatz (\ref{eq:fuzzyansatz}).
 
Returning then to our spherical fuzzy sphere ansatz for $\phi^i $, as argued in \cite{myers}, we can choose the generators to be the fundamental representation of the algebra
since this will correspond to the minimum energy configuration.
The physical radius of the fuzzy sphere is given by
\begin{equation}
r^2 = \frac{\lambda^2}{N} Tr(\phi^i \phi^j \delta_{ij}) = \lambda^2 R(t)^2 C,
\end{equation}
where $C$ is the quadratic Casimir of the representation defined by
\begin{equation}
\sum_i^3 (T^i)^2 = C 1_N,
\end{equation}
and $1_N$ is the $N \times N$ identity matrix. We also note that for the irreducible
representation, $C = N^2 -1 $, which can be approximated by $ N^2$ in the large $N$ limit.
In our analysis we will only be interested in this limit, as the case of finite $N$ has
additional complications due to the prescription of the symmetrized trace.
Combining all this information allows us to write the final form of the action as
\begin{equation}\label{eq:p=p'_action}
S= -\tau_{p'} \int d^{p'+1} \zeta N H^{-1} \sqrt{(1-H \lambda^2 \dot{R}^2 C)\\
(1+4\lambda^2 H C R^4)}- \tau_{p'} \int d^{p'} \zeta dt \frac{qN}{H}.
\end{equation}
Now, from the definition of the harmonic function, we see that the large $r$ limit corresponds to
Minkowski space, and the non-Abelian action reduces to the usual form for flat space \cite{myers, ramgoolam, costis}
We can now calculate the associated canonical momentum and energy density from the action, which are defined as follows
\begin{equation}
\tilde \Pi = \frac{\Pi}{\tau_p'} = N \lambda^2 \dot{R} C
\sqrt{\frac{(1+4\lambda^2HCR^4)}{(1- H \lambda^2 \dot{R}^2 C)}}
\end{equation}
\begin{equation}
\tilde E = \frac{E}{\tau_p'} = \frac{N}{H} \sqrt{\frac{(1+4\lambda^2 C H R^4)}
{(1-H\lambda^2 \dot{R}^2 C)}}-\frac{qN}{H},
\end{equation}
where the momentum is the derivative of the Lagrangian with respect to $\dot{R}$, and the energy is constructed
via Legendre transform. In addition we have divided out by a factor of $\int d^{p'} \zeta$ which loosely
corresponds to the 'volume' of each $Dp'$-brane.
To construct the potential energy we will find it useful to switch to the Hamiltonian formalism, where we write the energy in terms of the
conjugate variables.
\begin{equation}
\tilde E = \sqrt{N^2 H^{-2} (1+4\lambda^2 C H R^4) + \frac{\tilde \Pi^2}{H\lambda^2 C}}-\frac{qN}{H},
\end{equation}
which allows us to define the non-Abelian static potential via $V_{\rm eff}
=\tilde E(\tilde \Pi = 0)$.
\begin{equation}\label{eq:potential}
V_{\rm eff} = NH^{-1} \left(\sqrt{1+4 \lambda^2 C H R^4} - q \right),
\end{equation}
In order to consider the collapse of the fuzzy sphere, it will be more convenient to work in term of the physical radius $r$ rather than $R$. In
which case the potential can be written
\begin{equation}
V_{\rm eff} = N H^{-1} \left( \sqrt{1+\frac{4 H r^4}{\lambda^2 C}} -q \right),
\end{equation}
which is the gravitational potential generated by the background branes located at $r=0$.

It is useful to compare this result with that from the Abelian case, which was determined
to be \cite{branonium}
\begin{equation}
V^{abelian} = N\frac{(1-q)}{H},
\end{equation}
when we have $N$ probe branes separated by a distance larger than the string length.
Clearly we see that there is an additional term present arising from the non-Abelian nature of the effective action. Naively one might have
assumed that the potential for $N$ branes would be just $N$ times that for a single brane at lowest order. However, as we
can see there is an extra term corresponding to the additional
energy of the fuzzy sphere (or the vacuum energy of the non-commutative spacetime).
It is instructive to consider the behaviour of the potential in the different regions
of spacetime, but first we must ensure that there are no limiting constraints to be imposed on the
configuration.
Solving the energy equation for $\dot{r}$, we obtain the following equation of motion
which in turn will yield a constraint on the dynamics.
\begin{equation}\label{eq:radial_eom}
\dot{r}^2 = \frac{1}{H} \left( 1 - \frac{N^2}{(\tilde EH+qN)^2}
\left \lbrace 1 + \frac{4 H r^4}{\lambda^2 C} \right \rbrace \right).
\end{equation}
Since this equation is non-negative we see that the following constraint must
be satisfied, when we set the Chern-Simons part to zero,
\begin{displaymath}
1 \ge \frac{N^2}{\tilde E^2 H^2} \left \lbrace 1 + \frac{4Hr^4}{\lambda^2C} \right \rbrace.
\end{displaymath}
We consider what happens when we are in the near horizon geometry, as the constraint reduces to
the following expression
\begin{equation}\label{eq:throatconstraint}
1 \ge \frac{N^2}{\tilde E^2} \left(\frac{r^{7-p}}{k_p} \right)^2 \left \lbrace 1 + \frac{4k_pr^{p-3}}{\lambda^2 C}
\right \rbrace,
\end{equation}
For $p \ge 3$ the leading term in the expression is dominant and so we are effectively left with the following constraint
\begin{equation}
1 \ge \frac{N^2}{\tilde E^2} \left(\frac{r^{7-p}}{k_p}\right)^2.
\end{equation}
The supergravity solution implies that the term in parenthesis is already vanishingly small, which in turn
implies that the ratio $N/\tilde E$ can take a wide range of values and still satisfy this constraint.
We must emphasise at this point that the classical analysis
may break down as the fuzzy sphere collapses toward zero size, since the back reaction upon the source
branes will no longer be negligible and there will doubtless be correction terms to the energy
in this case which will invalidate this constraint. Furthermore there will also be the problem
of open string tachyon modes, which will arise as the branes approach distances comparable to the string length.
If we now consider the limiting case where $p<3$, the constraint equation becomes
\begin{equation}
1 \ge \frac{4}{\tilde E^2}\left(\frac{r^{7-p}}{k_p}\right)\frac{r^4}{\lambda^2},
\end{equation}
when we take the large $N$ limit. This solution has explicit dependence upon the ratio of the radius to the string length,
which we would expect to be larger than unity in order for us to have any faith in the effective field theory description.
This implies that the energy density can again be reasonably arbitrary as the supergravity constraint implies that the other
term is already vanishingly small. To be safe we will assume that $\tilde E >> N$ in what follows, as there is no ambiguity
in the constraints if this is fulfilled.
Interestingly if we reinstate the Chern-Simons contribution we find, to leading order, that the same constraints apply.

We now turn out attention to the large $r$ region ie flat space. In the Abelian case there are no constraints to be
imposed, and so the probe branes can move to an infinitely large distance from the
sources. In the non-Abelian case however, we can obtain an equation for the maximum radius of the fuzzy sphere
which can be written
\begin{equation}\label{eq:max_radius}
r^4_{max} = \frac{\lambda^2 C \tilde E^2}{4N^2} \left(1+\frac{2qN}{\tilde E}\right),
\end{equation}
from which we deduce that the orientation of the $Dp'$-branes plays the role of a small correction term provided we take our $\tilde E > N$ approximation.
This maximal distance represents the limit of our effective action, and it is likely that higher order correction
terms will allow us to consider limits such as $r_{max} \to \infty$.
We note, however, that this maximal distance is dependent upon the energy of the probe branes, and that by tuning
the energy we can effectively consider an unbounded solution in Minkowski space.
If we take the large $N$ limit and neglect the Chern-Simons part, this equation simplifies to
\begin{equation}\label{eq:r_max_approx}
r_{max} = \sqrt{ \frac{ \tilde E \lambda}{2}}\left(1+\frac{qN}{2\tilde E}+\ldots \right) \hspace{0.5cm} = \sqrt{\tilde E \pi l_s^2}\left(1+\frac{qN}{2\tilde E} +\ldots\right)
\end{equation}
which shows that the size of the fuzzy sphere is only dependent upon the energy of the solution.
This is what we expect from our knowledge of dielectric branes \cite{myers, hyakutake} and Giant Gravitons \cite{superstars, giant_gravitons}, which
are expanding brane solutions sourced by non-trivial background fields. Even though
we are only looking at a relatively simple example, we would expect to find some
similarities between these problems.

Armed with this our knowledge from the constraints we may proceed to investigate the behaviour of the 
effective potential. A quick calculation
shows that the potential has no turning point, therefore we shouldn't expect any stable bound states 
between the fuzzy sphere and the background branes.
It will be easier to analyse the behaviour of the solution in the two regions of spacetime, to learn more about the dynamics.
For vanishing $r$ we find the potential becomes
\begin{equation}
V_{\rm eff} \sim \frac{N r^{7-p}}{k_p} \left( \sqrt{1+\frac{4k_pr^{p-3}}{\lambda^2C}}-q \right).
\end{equation}
Now for $p \ge 3$, and sufficiently small radial distance, we may again ignore the radial contribution in the square root, provided that
\begin{equation}
r << \left(\frac{\lambda^2 C}{4k_p} \right)^{3-p} \nonumber,
\end{equation}
and we find this reduces to
\begin{equation}
V_{\rm eff} \sim \frac{N r^{7-p}}{k_p} (1-q),
\end{equation}
which we can see is identically zero if $q=1$, and is attractive if $q=-1$. This is
the same behaviour as seen for arbitrary $p$ in the Abelian case, and implies that the configuration can become
BPS at sufficiently small distances. However the size of this stabilisation radius is likely to be smaller than the 
string length, where our effective action is not valid.
Now if we consider
$p <3$ we find the potential is given by
\begin{equation}
V_{\rm eff} \sim \frac{N r^{7-p}}{k_p}\left( \sqrt{ \frac{4k_p}{\lambda^2 C r^{3-p}}} - q \right).
\end{equation}
which is attractive for all valid $p$ in this region. Therefore we see that to leading order, the probe branes are
always gravitationally attracted toward the source branes.

In the large $r$ limit, remembering that there is a maximum radius for the fuzzy sphere solution to hold, the
potential becomes.
\begin{equation}
V_{\rm eff} \sim N \left( \sqrt{\frac{4r^4}{\lambda^2 C}}-q \right),
\end{equation}
which we see will tend to a positive constant depending upon the exact size of the
maximum radius. If we substitute our solution (\ref{eq:max_radius}) into the potential, we find
\begin{equation}
V_{\rm eff} \sim \tilde E \sqrt{1+\frac{2Nq}{\tilde E}}-qN \sim \tilde E,
\end{equation}
where we have explicitly expanded out the square root term using our energy constraints. Thus the potential
energy is effectively the energy density at large $r$.
Before proceeding to solve (\ref{eq:radial_eom}), it is worth mentioning that the 'velocity' of the collapse
is a decreasing function of time. This is in stark contrast to the fuzzy sphere in a flat Minkowski background, where
we find that at small $r$, the velocity is a substantial fraction of the speed of light. The curved geometry of spacetime
in the near horizon limit acts in such a way as to slow the rate of collapse, in fact for an observer
on the background branes it would take an infinite amount of time for the sphere to reach zero size. Only if we switch to conformal
time will we see a finite time solution. This is an example of the usual red shift problem in 
General Relativity.

In the large $r$ region, we see that the harmonic function becomes unity and thus we would expect to find the usual
equations of motion for collapsing fuzzy spheres in flat space. Using the fact that the energy is conserved in time, we can integrate the
equation of motion to obtain the general form of the radial collapse in terms of Jacobi elliptic functions. By carefully selecting our initial
value of $r_0$ to be
\begin{equation}
r_0^4 = \frac{\lambda^2 C \tilde E}{4N} \left(\frac{\tilde E}{N} + 2q \right),
\end{equation} 
we find that the equation of motion is given by
\begin{equation}
r(t) = \pm r_0  {\rm JacobiCN} \left \lbrack 2\sqrt{\frac{2}{C}}\frac{r_0 t}{\sqrt{1+\frac{4r_0}{\lambda^2 C}}} ,\frac{1}{\sqrt{2}} \right \rbrack
\end{equation}
The form of this solution has been extensively discussed in \cite{ramgoolam, costis}, and so we will not say much about it here.
In this instance we know that the regime of validity for the solution is  $r^{7-p}>>k_p$ and so we find a simple monotonically expanding/contracting
solution without collapse toward  zero size. Thus the effective action should remain a valid description of the dynamics, and we do not have to worry about the physical
nature of the coordinate system being employed \cite{costis}. Interestingly this solution appears to be valid for arbitrary values of $p$ since all the $p$ dependence
arises in the form of the harmonic function, and gives rise to another example of the so called $p$-brane democracy.
The form of the equation of motion makes it difficult to obtain smooth analytic solutions interpolating between flat space and the near horizon
geometry. As a result we must regard the two regions as being distinct and choose boundary conditions such that it is possible to
match the solutions by hand. 

Turning our attention to the throat solutions, we see that the complicated form of the equation of motion makes analytic solutions
difficult to obtain. One case where we can make some progress is the $p=3$ background, as the 'fuzzy' term loses all radial dependence in this
instance. The solution is given in terms of a hypergeometric function, and it thus difficult to invert
\begin{equation}
t-t_0 \sim \pm \frac{\sqrt{k_3}}{r} \ _2 F_1 \left(\frac{1}{2}, \frac{-1}{8}, \frac{7}{8}, \frac{N^2 r^8}{\tilde E^2 k_3^2} \left\lbrace1+\frac{4k_3}{\lambda^2C} \right\rbrace \right).
\end{equation}
In the limit that the sphere collapses toward zero size, we can expand the hypergeometric function using the well known series expansion
\begin{equation}
t-t_0 \sim \pm \frac{\sqrt{k_3}}{r} \left(1-\frac{N^2 r^8}{14 \tilde E^2 k_3^2}\left\lbrace1+\frac{4k_3}{\lambda^2 C} \right\rbrace \right),
\end{equation}
which implies that at very late times the solution behaves as
\begin{equation}
r \sim \pm \frac{\sqrt{k_3}}{t-t_0}.
\end{equation}
The collapse of the sphere is described by the positive branch of the above solution, and is in fact an example of a simple power law solution. This power law
behaviour can be explicitly seen at late times by assuming that the dominant contribution to the denominator of (\ref{eq:radial_eom}) is unity. The 
resulting integral is trivial to perform and we obtain the general late time solution (dropping constants of integration)
\begin{equation}\label{eq:late_time_soln}
r \sim \pm \left(\frac{(p-5)(t-t_0)}{2\sqrt{k_p}} \right)^{2/(p-5)},
\end{equation}
the solution for $p=5$ must be calculated separately, but is simply proportional to an exponential
\begin{equation}
r \sim \exp\left(\pm\frac{t}{\sqrt{k_5}}\right).
\end{equation}
Thus we have shown that the solutions obey simple power law equations of motion as $r \to 0$. Of course, we must be careful in our interpretation
of these results as we expect correction terms to affect the validity of our effective action as the fuzzy sphere collapses.

We can solve the equations of motion numerically, which gives us some indication of the late time dynamics as measured by observers on the
source branes. For example, Figure 1 shows the numerical solution for $D0$ and $\bar{D}0$ branes. In order to generate this solution we
took $l_s =1$, $g_s=0.1$, $N=100$, $\tilde E=200$ and $M=1000$, whilst retaining the full form of the harmonic function but taking the large $N$ limit.
Although the parameter space of solutions is large, we expect the numerical solutions to be representative of more general behaviour. In fact 
we investigated the dynamics for various ranges of energy, and found approximately the same solutions with all the solution curves collapsing onto
one another at very small distances.
The analytic solution clearly shows that the radius collapses rapidly when the
metric is approximately flat, but decelerates as the sphere enters the near horizon geometry. We expect that our solutions will break down as the
probes near the source branes, although it is useful to recall that $D0$-branes can probe distances smaller than the string length and so
the solution may be valid for some time. The plot shows that the brane and anti-brane follow similar trajectories as they cross
into the near horizon region and are thus indistinguishable. Our analysis of the potential suggests that it should vanish for the $D0$-brane solution
as $r\to 0$. Clearly our plot shows that this must happen at a distance smaller than the string scale. 

Figure 2 shows the solutions for the $D4$ and $D5$-brane backgrounds using the same parameters, but ignoring the
Chern-Simons term. The five brane solution indeed tends toward an exponential at late times as expected from our simplified
analytic solution.

Figure 3 shows the solution for the $D3$ and $\bar{D}3$-branes. In this instance we can clearly see that the fuzzy sphere associated
with the $D3$ solution collapses faster than the $\bar{D}3$ solution when they are in flat space. This is because the $D3$-branes are
more strongly attracted to the sources than the $\bar{D}3$-branes.
However as they cross into the near horizon geometry, both spheres tend to the same radius as the Wess-Zumino term becomes negligible which
accounts for the similarity in their dynamics.

\begin{figure}
\begin{center}
\epsfig{file=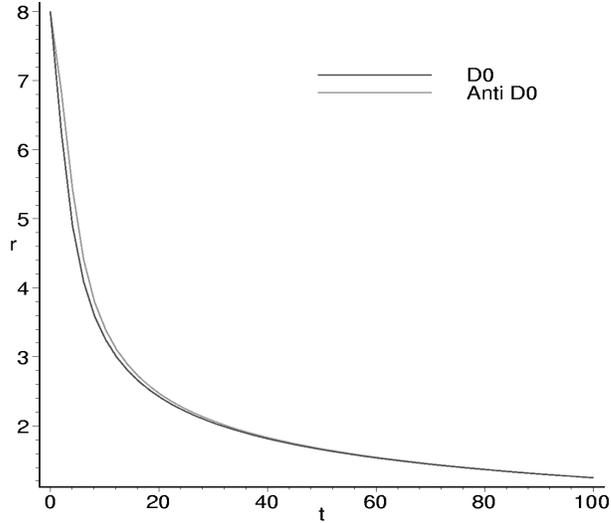, width=8cm,height=8cm}
\caption{Numerical solution to the equations of motion for the $D0$-brane background.}
\end{center}
\end{figure}

\begin{figure}
\begin{center}
\epsfig{file=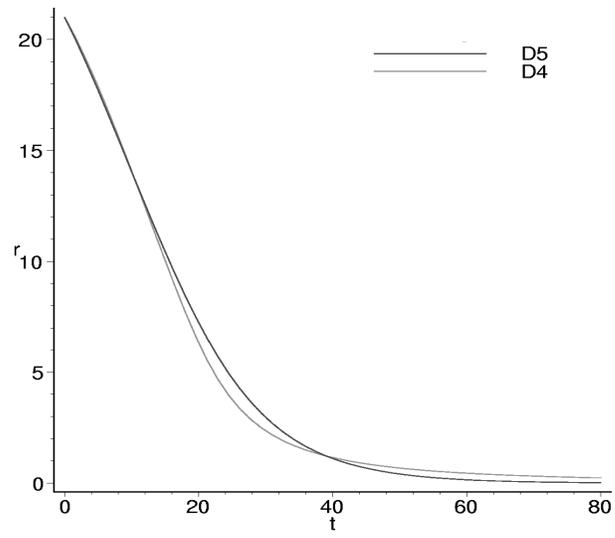, width=8cm,height=8cm}
\caption{Solutions for $D4$ and $D5$ brane backgrounds, ignoring the Chern-Simons coupling.}
\end{center}
\end{figure}

\begin{figure}
\begin{center}
\epsfig{file=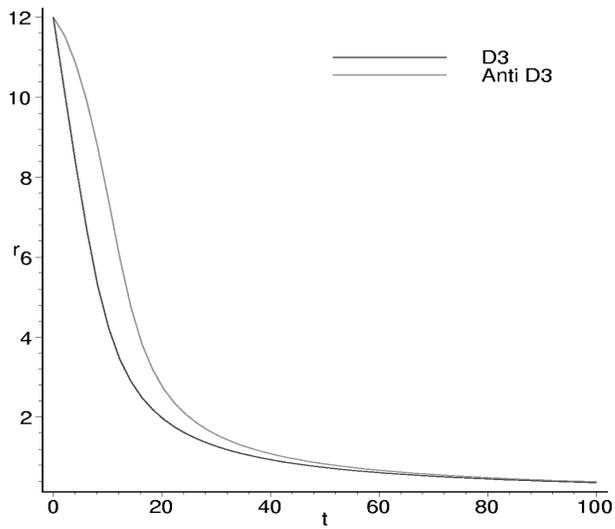, width=8cm,height=8cm}
\caption{Solutions for the fuzzy sphere sourced by $D3$ and anti $D3$-branes.}
\end{center}
\end{figure}

The difficulty in analytically solving the integral equation of motion is related to the fact that it describes curves on hyper-elliptic Riemann surfaces,
with the infinitesimal time playing the role of a holomorphic differential. The velocity and the radius can each be regarded
as two complex variables related by a single constraint. We can use the simplified Riemann Hurwitz formula to calculate the genus, g, of
the underlying surface
\begin{equation}
g = \frac{1}{2}(B-2),
\end{equation}
where $B$ refers to the number of branch points of our solution. It is fairly straight forward to see that the $p=6$ and $p=5$ cases
correspond to genus 2 surfaces, $p=3, 4$ give rise to genus 3 surfaces, $p=2, 1$ are genus 5 surfaces and $p=0$ defines a genus 7 surface.
Thus as we decrease the dimensionality of the background branes, we find surfaces of higher and higher genus. Obviously this leads to
the difficulty in obtaining an analytic solution to the equation of motion. Even if we include the Chen-Simons term in the equation of motion,
this doesn't modify the number of branch points.
As in \cite{costis} it may be possible to reduce the integral for the genus 3 and 5 surfaces into integrals over products of genus 1 surfaces using the special
symmetries present. The solution in flat space corresponds to a genus 1 surface, which is why we find an explicit solution to the equation of motion.
This suggests that the Riemann surface describing the curved backgrounds is actually of high genus, with the branch points on the complex plane being
totally unresolvable when the cycles are large.

\subsection{Dynamics in the $p \ne p'$ case.}
We now turn our attention to the more general case where $p \ne p'$. However, as we
are only looking at the leading order terms in the action we find that there
is no Chern-Simons term except for the $p=6, p'=0$ case. But for the purpose of this note,
we will neglect this contribution.
The action in this instance is a simple extension of (\ref{eq:p=p'_action}) and can be written as
\begin{equation}
S=-\tau_{p'} \int d^{p'+1}\zeta N H^{(p-p'-4)/4} \sqrt{(1+4H\lambda^2CR^4)(1-H\lambda^2C\dot{R}^2)},
\end{equation}
which clearly reduces to the expression in the previous section when taking the $p=p'$ limit.
We will again divide out by the 'mass' of the brane to find a closed expression for the canonical momentum , which
turns out to be
\begin{equation}
\tilde \Pi = NH^{(p-p'-4)/4} \lambda^2 C \dot{R} \sqrt{\frac{1+4H\lambda^2CR^4}{1-H\lambda^2C\dot{R}^2}},
\end{equation}
and the corresponding energy is obtained via Legendre transformation in the usual manner.
\begin{eqnarray}
\tilde E &=& NH^{(p-p'-4)/4} \sqrt{\frac{1+4H\lambda^2CR^4}{1-H\lambda^2C\dot{R}^2}} \\
&=& \sqrt{N^2H^{(p-p'-4)/2}(1+4H\lambda^2CR^4)+\frac{\tilde \Pi^2  }{H\lambda^2C}}, \nonumber
\end{eqnarray}
Following results from the previous section we define the effective potential to be
\begin{equation}
V_{eff} = N H^{(p-p'-4)/4}\sqrt{1+\frac{4Hr^4}{\lambda^2C}},
\end{equation}
which is clearly the general extension of (\ref{eq:potential}) when there is no Chern-Simons coupling term. Interestingly the extra
energy due to the fuzzy sphere actually breaks the supersymmetry in this case. 
Using the conservation of energy we also have a modified constraint condition
\begin{equation}
1 \ge \frac{N^2 H^{(p-p'-4)/2}}{\tilde E^2}\left(1+\frac{4Hr^4}{\lambda^2C} \right).
\end{equation}
In the near horizon geometry we see that the RHS blows up as as the radius tends to zero when $p-p'>4$ which, because of the dimensionality of the branes,
 implies that for the $p=6, p'=0$ case the energy must go to infinity as
the fuzzy sphere collapses in order to satisfy the constraint. All of the other solutions are satisfied
for arbitrary energy in this limit. This tells us that the $D6-D0$ solution will not collapse to zero size, instead it
will be energetically favourable for the fuzzy sphere to expand in the near horizon geometry.
In the large $r$ limit we again expect there to be a maximum size for the fuzzy sphere solution, which is
given by (\ref{eq:r_max_approx}) when we take the large $N$ limit.

By analysing the behaviour of the effective potential, we should get a general understanding of the dynamics of the fuzzy sphere
as the probe branes are attracted to the source branes. In general we see that the potential is always attractive, implying
that the fuzzy sphere will eventually collapse down toward zero size. The cases where this isn't true are for $p=6, p'=0$
which has a repulsive potential at small radius \cite{susskind}, exactly as we would expect from energy considerations. We will
have more to say about the $D6-D0$ configuration in a later section as we expected it to be related to the non-Abelian extension
of the Quantum Hall soliton. The other case where the potential does not vanish is when $p-p'=4$, corresponding to the cases
$p=6,p'=2$; $p=5,p'=1$ and $p=4,p'=0$. In these cases we see that the potential tends to $N$ with vanishing radius. Again this
should be expected as the branes are all parallel and this is precisely the supersymmetry preserving condition in the Abelian theory
\cite{branonium}, however this may well occur at distances beyond the regime of validity of our effective theory.

Solving the equations of motion in the general case is far from trivial, as the integral equation describes surfaces of varying genus.
For completeness we have written the genus associated with all the possible values of $p, p'$ in our analysis. Note that as the factor $p-p'$
increases, the genus of the surface associated with the solution decreases. For example in the $p-p'=4 $ case (not including $p=6$), we
see that the Riemann surface becomes a simple two-sphere. This is interesting as we know that this is exactly the supersymmetry
preserving condition in the Abelian theory \cite{branonium}, and a quick calculation verifies that the Abelian equation also yields
a genus 0 surface even in the $p=6, p'=2$ case. This poses the question of whether there is some deeper connection between the
preservation of supersymmetry and the underlying Riemannian geometry.
An example solution can be found in the $p=4, p'=0$ case which will be valid 
when $r$ satisfies the following constraint, $\lambda^2 \tilde E^2 >> 4k_4 r$. 
Upon integration we find
\begin{equation}
r \sim r_0 \pm \frac{4 \tilde E k_4}{(\tilde E^2 -N^2)t^2},
\end{equation}
where we must take the negative branch of the solution to approximate the collapsing fuzzy sphere.

\begin{center}
\begin{tabular}{|c|c|c|c|c|c|c|c|c|c|c|c|c|c|c|c|c|}
\hline
$p$&  6& & & &5& & &4& & &3& &2& &1&0\\
\hline
$p'$&  6&4&2&0&5&3&1&4&2&0&3&1&2&0&1&0\\
\hline
genus & 2&2&1&1&2&1&0&3&2&0&3&1&5&2&5&7\\
\hline
\end{tabular}
\end{center}
\subsection{Corrections from the symmetrized trace.}
In our work so far we have only considered the leading order Lagrangian,
and neglected any $1/N$ corrections. However, these terms can be calculated allowing
corrections to the effective potential to be found.
We remind the reader that to lowest order, we have calculated the energy density to be
\begin{displaymath}
\tilde E = \frac{\delta \mathcal{L}}{\delta \dot{R}}\dot{R}-\mathcal{L}.
\end{displaymath}
Based upon arguments in \cite{ramgoolam} we know that the corrections to next order are
given by
\begin{equation}
\tilde E_1 = \left( 1-\frac{2}{3}C \frac{\partial^2}{\partial C^2} \right) \tilde E,
\end{equation}
where we have dropped all the Chern-Simons terms to make things clearer.
We differentiate our expression for $\tilde E$ in order to find the next order corrections
to the effective potential. Note that for static BPS configurations such as the $D1-D3$ intersection,
all the symmetrized trace correction terms are zero. We don't anticipate the same situation occurring here
because the Chern-Simons coupling is independent of $C$ and will drop out when we differentiate the Lagrangian. Since it
is this coupling which (in the Abelian case at least) preserves the bulk supersymmetries, we expect
that higher order corrections will not be BPS configurations, and so we will find non-zero correction terms
to all orders. Our calculation for the general case, gives us the first order correction to the potential
\begin{equation}
\Delta V_{\rm eff} = \frac{8NH^{(p-p'+4)/4}r^8}{3 \lambda^4 C^3 \left( 1+\frac{4Hr^4}{\lambda^2 C} \right)^{3/2}},
\end{equation}
where we have made use of the near horizon approximation. Once more we find that the solution depends heavily upon
the dimensionality of the branes involved. Firstly, we consider the case
when $p \ge 3$. In this instance the correction term becomes;
\begin{equation}
\Delta V \sim \frac{8Nr^8b}{3 \lambda^4 C^3}\left(\frac{1}{r^{7-p}}\right)^{(p-p'+4)/4}.
\end{equation}
Where we have introduced $b = k^{(p-p'+4)/4}$ for simplicity. In general the factor of $p-p'$ can
only take the integral values of $6, 4, 2$ or $0$, and so it is easily
noted that the potential tends to zero as $r\to 0$ for all values of $p$ and $p'$
in this particular range. If we move on to consider the case where $p<3$ then
$p-p'$ is limited to be either $2$ or $0$. The correction term in this instance reduces to
\begin{equation}
\Delta V \sim \frac{8Nr^8b}{3 \lambda^4 C^3}\left( \frac{1}{r^{7-p}}\right)^{(p-p'+4)/4} \left( \frac{\lambda^2C}{4k_pr^{p-3}}\right)^{3/2}
\end{equation}
This potential again tends to zero with $r$ for all values of $p$ and $p'$, which is in agreement with 
our general expectations from the behaviour of the leading order term. Thus the correction doesn't alter the overall dynamics of the fuzzy sphere, 
and we don't find any bounce solutions.  However it should be noted that if we relax our throat approximation and look at large $r$, we would expect to find
differing behaviour. For example \cite{ramgoolam} showed that there are bounce solutions for the $D0$-solution in flat Minkowski space when
the $1/N$ sub leading order terms are taken into account. It is well known that $D0$-branes may probe distances much smaller than the string length \cite{douglas}, however
the curved backgrounds we have been studying in this section appear to impose constraints upon this behaviour. 
\subsection{Remarks on the $D6$-$D0$ solution.}
In this section we will briefly comment on the $p=6, p'=0$ solution as there is a similarity with
the Quantum Hall Soliton (QHS), which we will briefly introduce below.

The stringy QHS was introduced \cite{susskind} as a way of establishing the link between condensed
matter physics and string theory. To construct the QHS, we imagine a background of $k$
coincident $D6$-branes with $k$ strings emerging from them. The transverse space can
be parameterised simply by $\mathbb{R} \times \mathbb{S}^2$, and we wrap a $D2$-brane over the
$\mathbb{S}^2$. However it is known that this configuration is unstable, and so we are forced to introduce
$N D0$-branes, which are dissolved into the $D2$-brane world volume. Since it is well known 
that $D6$ and $D0$-branes repel each other (due to the energy becoming infinite at small distances), this stabilises the QHS.
The world volume of the spherical $D2$-brane, in this instance, becomes the surface where the quantum hall fluid lives.

This is a purely Abelian theory in terms of the $D2$ picture, however our Non-Abelian 
construction can provide information on the dual picture. This is because we can consider
$N D0$-branes in the supergravity background of $M$ coincident $D6$-branes. We expect that 
the fuzzy sphere ansatz will play the role of the $D2$-brane with flux on the Abelian side, furthermore
we anticipate that the $D0$-branes can be regarded as being the endpoints of fundamental strings which
start on the background $D6$-branes. The only difference is that we are neglecting the open string contributions
from the background branes to the probe branes.

We have already seen that the effective potential for this (bosonic) configuration can be written as
\begin{equation}
V = N \sqrt{H}\sqrt{1+\frac{4Hr^4}{\lambda^2 c}},
\end{equation}
where the harmonic function, $H$, can be approximated in the near horizon limit by
\begin{equation}
H \approx \frac{Mg_sl_s}{2r}. \nonumber
\end{equation}
We now determine, by differentiating the potential, that there is a minimum at the distance
\begin{equation}
r_{min} = \left( \frac{\pi^2 l_s^3 N^2}{Mg_s} \right) ^{1/3},
\end{equation}
where we have explicitly employed the use of the large $N$ limit. This is exactly the same result that was obtained for the stability
of the spherical $D2$-brane with flux in terms of the gravitational Myers effect effect \cite{gravitational_dielectric}.
We wish to compare this result to the one calculated in \cite{susskind}. In that paper they used a coordinate
rescaling to simplify the initial background metric. The scaling is given by
\begin{displaymath}
r= \rho \left(\frac{Mg_s}{2} \right)^{-1/3},
\end{displaymath}
and consequently the equation for the stabilisation radius is given by
\begin{equation}
\rho_* = \frac{ (N\pi)^{2/3} l_s}{2}.
\end{equation}
Performing the same rescaling in our Non-Abelian dual picture gives the result
\begin{equation}
\rho_* = \frac{(N\pi)^{2/3}l_s}{2^{1/3}}
\end{equation}
which is almost identical to the Abelian theory. In fact the discrepancy between the two radii is due to the contribution
from the $k$ strings on the Abelian side, which has been neglected in our analysis. 
In fact the string contribution alters the stabilization radius
by a factor of $2^{-2/3}$.
If we reconstruct the QHS, but neglect the stringy contribution and allow for time dependent radial solutions we obtain the following action
\begin{equation}
S = -\tau_2 \int d^3 \zeta Hr^2 \sin(\theta) \sqrt{(1-H\dot{r}^2)\left(1+\frac{\lambda^2N^2}{4Hr^4}\right)},
\end{equation}
where we use the usual spherical coordinates on the $D2$-brane worldvolume and the flux on the brane is given by
\begin{equation}
F_{\theta \phi} = \frac{N\sin(\theta)}{2},
\end{equation}
which satisfies the usual quantization conditions. For a more rigorous explanation of the derivation we refer the reader
to \cite{susskind} for more details. We can integrate out the angular dependence to find an exact expression for
the Lagrangian
\begin{equation}
\mathcal{L}=-\tau_2 4\pi r^2 H \sqrt{(1-H\dot{r}^2)\left(1+\frac{\lambda^2 N^2}{4Hr^4} \right)}.
\end{equation}
Using this we can easily construct the static potential for the Abelian theory in the near horizon geometry, which we find to be
\begin{equation}
V = \frac{kr}{\lambda}\sqrt{1+\frac{\lambda^2 N^2}{2kg_sl_sr^3}}.
\end{equation}
Although this appears to be different from the non-Abelian potential, they are in fact identical as can be verified with
a simple expansion. Thus the theories are in fact dual to one another, which we can further exhibit by analysing the equations of motion
for the radion fields. Using subscripts $A$ and $N$ to represent the two theories, we find the result
\begin{eqnarray}
\dot{r}_A^2 &=& \frac{1}{H}\left(1-\frac{16 \pi^2 \tau_2^2 H^2 r^4}{E^2}\left\lbrace 1+\frac{\lambda^2 N^2}{4Hr^4} \right\rbrace \right), \\
\dot{r}_N^2 &=& \frac{1}{H}\left(1-\frac{\tau_0^2 N^2 H}{E^2}\left\lbrace1+\frac{4Hr^4}{\lambda^2 C} \right\rbrace \right) \nonumber.
\end{eqnarray}
If we take the large $N$ limit and carefully expand these equations we see that they are identical. This was noted \cite{ramgoolam}
for the case of a fuzzy sphere in flat space, and as expected this duality continues to hold in a curved geometry.
On the Abelian side we find an explicit example of the gravitational dielectric effect, whilst on the non-Abelian side we have the 
gravitational Myers effect. It would be useful to include the terms coming from the strings in our work, as this would 
be the dual of the QHS, however this is expected to be complicated as the strings are charged under $U(M)$ on one end and $U(N)$ on the 
other. The corresponding trace over the Chan-Paton factors will be expected to yield an extra term in the DBI forcing the fuzzy sphere
to stabilise at a smaller radius due to the tension of the strings.
As a further remark, we should note that this duality only holds for the $p=6, p'=0$ case. We could
consider a different background source such as $D4$, $D2$ or $D0$-branes, with the $D2$-wrapped over a transverse $\mathbb{S}^2$ whilst the 
remaining transverse coordinated are set to zero. Unfortunately the corresponding solutions do not map across to the non-Abelian construction
where we would have $D0$-branes probing each of these background solutions. This is because we are losing information about the theory
by setting some of the Abelian degrees of freedom to zero.

It is interesting to examine the stability of our solution with regards to $D0$-brane emission. It was argued for the QHS that there is an energy barrier proportional to $N$, 
preventing the tunnelling of $D0$-branes out of the $D2$ brane. In fact it requires energy to be out into the system to
remove the $D0$-brane. Therefore the QHS appears to be stable with respect to particle emission \footnote{ \cite{susskind} also noted that there could be possible nucleation of the $D2$-brane
causing another $D2$ brane to appear inside the original one. Although we can consider multiple fuzzy spheres by selecting an ansatz which is a reducible representation,
this does not correspond to the picture on the Abelian side. It would be certainly interesting to consider a non-Abelian description of this.}.

The potential at the stable radius in our dual picture can be written explicitly as
\begin{equation}
V = N\sqrt{\frac{(Mg_s)^{4/3}}{2(N\pi)^{2/3}}}\sqrt{1+\frac{N^2}{2C}},
\end{equation}
where we are using the dimensionless potential obtained from $\tilde E$. We now revert to proper time as measured by 
an observer on the fuzzy sphere, which allows us to re-write the minimised potential with respect to proper time
\begin{equation}
V_T(N) = \sqrt{\frac{N}{\pi}}\frac{Mg_s}{2^{3/4}}\sqrt{1+\frac{N^2}{2C}}.
\end{equation}
Now imagine that the soliton emits a single $D0$-brane into the bulk, the change in the potential - to leading order in $1/N$, and taking the large $N$ limit - can be
approximated by
\begin{equation}
V_T(N) - V_T(N-1) \sim \sqrt{\frac{3}{N\pi}} Mg_s.
\end{equation}
We now need to compare this with the potential energy of a single $D0$-brane attached to a fuzzy sphere located at the stabilisation
radius. Although our effective  action is valid as a large $N$ expansion, we can use it to determine the potential for a single brane
provided that we neglect the back reaction terms between brane and fuzzy sphere. By adding this contribution to the one
calculated in the previous line we see that
\begin{equation}
V_{tot} \sim \frac{Mg_s}{\sqrt{\pi}}\left(\sqrt{\frac{3}{N}}+\frac{1}{\sqrt{2}} \right),
\end{equation}
which is larger than the potential of the stable fuzzy sphere. Thus we conclude that the solution appears to be stable with regard
to emission. This gives us an estimate of the binding energy of the $D0$-branes in the near horizon region, which we interpret as the 
energy barrier needed for quantum tunnelling
\begin{equation}
E_{\rm binding} \sim \nu g_s \sqrt{N},
\end{equation}
where we have made use of  the ratio $\nu =M/N$ to simplify the result. In the QHS picture this corresponds to the definition of the filling ratio. Clearly the barrier is proportional to $N$, thus in
the large $N$ limit we would expect the fuzzy sphere to be stable.

The supergravity picture of this case is then the following. If the fuzzy sphere is initially large, then the metric is 
approximately Minkowski and we have our usual collapsing solution \cite{ramgoolam} with velocity approaching that of light. As the
$D0$-branes enter the near horizon geometry they decelerate (from the $D6$ viewpoint) until they oscillate around the minimum of
the potential, eventually forming a bound state at $r_{min}$.
If on the other hand, the fuzzy sphere is initially small, then the gravitational dielectric effect forces the configuration to
expand until it reaches the stabilisation radius - at which point it settles into its bound state after oscillation.
\section{Inclusion of Angular Momentum.}
In the Abelian case, the inclusion of angular momentum terms in the action is trivial since
all the coordinates commute. This will clearly not be the case in the Non-Abelian version, and so
we must choose a specific ansatz. A fuzzy cylinder ansatz was introduced in \cite{fuzzy_cylinder}, which was able to rotate about three 
independent axes. However, this ansatz proves to be restrictive on the dimensionality of the background brane solutions
limiting them to $p \le 3$, although it may be useful in describing dual versions of supertubes \cite{supertubes} and we will have a closer look at it in the
next section.
Instead we choose a different ansatz corresponding to rotation in the $\phi^6-\phi^7$ plane, 
\begin{eqnarray}
\phi^6 &=& R(t) \cos(\theta) T_3, \nonumber \\
\phi^7 &=& R(t) \sin(\theta) T_3, \nonumber \\
\phi^8 &=& R(t) T_1, \nonumber \\
\phi^9 &=& R(t) T_2.
\end{eqnarray}
This means that the resulting action will only be valid for $p<6$, and so we will not be able to
consider rotation in the gravitational Myers effect picture. The action for this particular ansatz
can be calculated, and we find 
\begin{equation}
S=-\tau_{p'} \int d^{p'+1} \zeta \sum_{j=0}^N NH^{(p-p'-4)/4} \sqrt{(1+4H\lambda^2CR^4)(1-H\lambda^2C \dot{R}^2-H\lambda^2R^2\dot{\theta}^2 \lambda_j^2)}.
\end{equation}
where $\lambda_j$ is the $j$th eigenvalue of the matrix $(T_3)^2$ (using a matrix representation for the diagonal generator). If we expand the action out to leading order
this enables us to isolate the $\lambda_j$ dependence and we can perform the sum to obtain
\begin{equation}
\sum_{j=0}^N \lambda_j^2 = \frac{N}{12}(N^2-1) = \frac{CN}{12}.
\end{equation}
In general, the inclusion of angular momentum for the fuzzy sphere is non-trivial. If we employ a convention where the subscript on the
$\lambda$ implies summation over that variable then we find the exact solution for the static potential in physical radius is given by
\begin{equation}
V_{eff} = \frac{NH^{(p-p'-4)/4}}{\sqrt{1-H\lambda^2 R^2\dot{\theta}^2 \lambda_{j}^2}}\sqrt{1+\frac{4Hr^4}{\lambda^2C}}\left(\frac{HNr^2\dot{\theta}^2}{12}+\sqrt{1-H\lambda^2R^2\dot{\theta}^2 \lambda_{k}^2}
\sqrt{1-H\lambda^2 R^2\dot{\theta}^2 \lambda_{j}^2} \right),
\end{equation}
where $\dot{\theta}$ corresponds to the angular velocity of the fuzzy sphere. By setting this term to zero we recover
the result for the purely radial collapse, as we would anticipate. Even though we cannot find a closed form solution for the potential we can still make some 
comments about the dynamics of the fuzzy sphere. Interestingly we expect that the potential will vanish in the $r \to 0$ limit, as the only case
where there is the possibility of a bound state is when $p-p' > 4$ corresponding to the $p=6, p'=0$ case we investigated in the previous section. Unfortunately
our choice of ansatz doesn't allow for this to be investigated here. This tells us that the angular momentum term cannot counteract the gravitational force
exerted by the source branes, and the fuzzy sphere will always collapse.

\subsection{Alternative ansatz.}
Thus far our analysis has been exact but not concise, so it is useful to consider an alternative
ansatz which allows us to incorporate angular momentum in a clear manner. Since we require two
transverse scalars to define a plane in the transverse space, and at most each plane is
parameterised by one of the generators of the representation, we are led to the conclusion that we
need six transverse scalars to introduce angular momentum. This will place severe
restriction upon the dimensionality of the branes that we can consider in our solution. In fact we find
that at most we can consider a $D3$-brane background.
We choose to parameterise the six transverse scalars as follows:
\begin{eqnarray}
\phi^1 &=& R(t) \rm cos(\theta) \hspace{0.5cm} \phi^2 = R(t) \rm sin(\theta) \nonumber \\
\phi^3 &=& R(t) \rm cos(\theta) \hspace{0.5cm} \phi^4 = R(t) \rm sin(\theta) \nonumber \\
\phi^5 &=& R(t) \rm cos(\theta) \hspace{0.5cm} \phi^6 = R(t) \rm sin(\theta)
\end{eqnarray}
Thus we are breaking the $SO(6)$ symmetry of the transverse space to $SO(2) \times SO(2) \times
SO(2)$, and choosing the same angle $\theta$ to parameterise the three planes.
This may seem a rather restrictive ansatz, but it will actually allow us to make some
progress.
The action in this case becomes
\begin{equation}
S=-\tau_{p'} \int d^{p'+1} \zeta STr \left( H^{(p-p'-4)/4}\sqrt{1-H\lambda^2 C
(\dot{R}^2+R^2 \dot{\theta}^2))(1+4\lambda^2 H R^4C)}\right),
\end{equation}
with a possible Chern-Simons term, defined up to a constant factor
\begin{equation}
S_{CS}= -\tau_{p'}\delta_{p'}^p \int dt \frac{q}{H}.
\end{equation}
Since both terms in the
Born-Infeld part of the action are proportional to the identity matrix, 
we find that the $STr$ reduces to $Tr$ to leading order in large $N$.
Finally we obtain
\begin{equation}
S=-\tau_{p'} \int d^{p'+1} \zeta NH^{(p-p'-4)/4}\sqrt{(1-H\lambda^2 C
(\dot{R}^2+R^2 \dot{\theta}^2))(1+4\lambda^2 H R^4C)}.
\end{equation}

We can now proceed as usual by switching to the Hamiltonian formalism and writing the
canonical energy density as
\begin{equation}
\tilde E = \sqrt{ N^2 H^{(p-p'-4)/2} (1+4\lambda^2 CHR^4)
+\frac{1}{H\lambda^2 C} \left(\tilde \Pi^2 + \frac{\tilde L^2}{R^2}\right)}.
\end{equation}
Switching to the physical radius $r$, we find that the effective potential becomes
\begin{equation}
V_{\rm eff} = \sqrt{N^2 H^{(p-p'-4)/2} \left( 1+\frac{4Hr^4}{\lambda^2 C} \right) + \frac{\tilde L^2}{Hr^2}}
\end{equation}
Where we must remember that this equation is only valid for $p \le 3$, and the energy density and the angular
momentum are the conserved charges If we set the
angular momentum term to zero we recover the potential for a radially collapsing solution, as 
we would expect. For ease of calculation we choose to rescale the potential by a factor of $N$. This
is possible because there is an $N^2$ term in the angular momentum density. The 
resulting non-Abelian and Abelian potentials are written below for comparative purposes
\begin{eqnarray}
\bar{V}_{\rm eff} &=& \sqrt{H^{(p-p'-4)/2} \left( 1+ \frac{4Hr^4}{\lambda^2C} \right) + 
\frac{\tilde L^2}{Hr^2}}, \\
V^{\rm abelian} &=& \sqrt{H^{(p-p'-4)/2} + \frac{\tilde L^2}{H r^2}}. \nonumber 
\end{eqnarray}
Simple analysis of the potential in the non-Abelian case shows that it is a monotonically decreasing function
for all valid $p$ and $p'$ in this region. Therefore there is no possibility of the formation of bound states,
in the same way that there are no bound orbits in the Abelian theory \cite{branonium}.
Once again it is useful to look at the equations of motion to determine if there
are any constraints to be imposed on the solution. We wish to consider a case
where the energy density and the angular momentum density are constant. Thus,
we find the following expression
\begin{equation}
\dot{r}^2 = \frac{1}{H} \left( 1 - \frac{1}{\tilde E^2}\left\lbrack N^2 H^{(p-p'-4)/2}\left\lbrace1+\frac{4Hr^4}{\lambda^2 C} \right\rbrace + \frac{\tilde L^2}{Hr^2} \right\rbrack \right).
\end{equation}

If we assume that the angular momentum takes some fixed, non-zero value - then we can consider how the constraint equation is modified
in the asymptotic limit of $r \to 0$
\begin{equation}
1 \ge \frac{1}{\tilde E^2} \left( \frac{4N^2 k_p^{(p-p'-4)/4} k_p}{\lambda^2 C r^{((7-p)(p-p'-4)+6-2p)/2}} + \frac{\tilde L^2r^{5-p}}{k_p} \right).
\end{equation}
This appears to have a complicated dependence upon $r$, however because of the restrictions from the ansatz we know that there are only two
possible cases we can consider, $i)$ $p-p'=2$ and $ii)$ $p-p'=0$.
The first case reduces the constraint to the following
\begin{equation}
1 \ge \frac{1}{\tilde E^2} \left( r^4 + \tilde L^2 r^{5-p} \right).
\end{equation}
It is clear that as $r$ vanishes the contribution from the angular momentum term also vanishes and the energy density can be relatively 
arbitrary, as already discussed.
The second condition implies a similar result, however the dimensionalities of the branes involved plays a role in determining
how quickly the lead term vanishes.
\section{Non-BPS branes.}
It is well known that BPS branes are soliton solutions of Non-BPS branes, so it is natural to 
enquire about the dynamics of these branes in various backgrounds. In this section we will look
at the action for $N$ Non-BPS branes in the $Dp$-brane background and try and study the dynamical
evolution of the fuzzy sphere in this instance. This will not be as straight forward to analyse
as the BPS case \cite{non_bps_dynamics}, as there is the additional complication of open string tachyon modes condensing on the world volume.
We start with the generalised non-Abelian action for the probes, which can be expanded to lowest order
 \cite{non_abelian_non_bps}.
\begin{equation}
S=-\tau_{p'} \int d^{p'+1} x Str V(T) H^{(p-p'-4)/4} \sqrt{(1-\lambda^2 H \dot{\phi}^2 - H^{1/2}\lambda \dot{T}^2)(1-\frac{\lambda^2 H }{2}[\phi^i,\phi^j][\phi^j,\phi^i])}.
\end{equation}
The tachyon field is dimensionless and we are assuming, like the transverse scalars, that it is purely time dependent. This also ensures that the Chern Simons term
vanishes to lowest order when we use the static gauge. $V(T)$ is the potential for the tachyon field, which describes the changing tension of each 
of the branes. Note that in this section we will be using the standard form of the tachyon potential 
\cite{non_bps_action, tachyon_transform, sen, time_dependence, non_bps_dynamics} where $V(T) \propto 1/\cosh(T)$.
It would be certainly be interesting to study the case of spatially dependent tachyon fields, as their
classical solutions give rise to kink-antikink solutions on the world volume \cite{sen}. We now make use of the $SU(2)$ ansatz, $\phi^i = R(T) T^i$ as usual
and find that the action reduces to the form
\begin{equation}
S=-\tau_{p'} \int d^{p'+1} x N V(T) H^{(p-p'-4)/4} \sqrt{(1-\lambda^2 CH \dot{R}^2-\lambda H^{1/2} \dot{T}^2)(1+4H\lambda^2 CR^4)},
\end{equation}
where we have performed the symmetrized trace to bring the Casimir into the action. As it stands this is perfectly
acceptable for us to analyse the dynamics. However the presence of the tachyon makes things difficult since it
will not decouple from the equation of motion for the radion field. It is more useful to modify this action
to another equivalent form, and investigate the dynamics by finding another conserved charge.
In order to do this, we choose to rescale the tachyon field \cite{tachyon_transform, non_bps_dynamics}
\begin{equation}
\frac{\tilde T}{\sqrt{2}}= \sinh \left(\frac{T}{\sqrt{2}}\right),
\end{equation}
which transforms the action into
\begin{equation}
S=-\tau_{p'} \int d^{p'+1}x \frac{N H^{(p-p'-4)/4}}{\sqrt{F}}\sqrt{1+\frac{4Hr^4}{\lambda^2C}}\sqrt{1-H \dot{r}^2-\frac{H^{1/2}\lambda \dot{T}^2}{F}}.
\end{equation}
Where $F$ now controls the behaviour of the tachyon and the changing tension of the probe branes, which is simply
\begin{equation}
F(T)=1+\frac{T^2}{2},
\end{equation}
and we have also chosen to write the new tachyon field in terms of $T$ for ease of notation.
This form of the action allows us to investigate the dynamics of the Non-BPS brane when the tachyon
field is large \cite{non_bps_dynamics}. At this juncture we must point out that there may be objections to
using this form of rescaling, as we are assuming that it will hold true in a gravitating 
background. It is well known that there are many effective descriptions for the tachyon field,
with each one defined on a specific section of tachyon moduli space. However as there has been little
progress in constructing non-Abelian versions of these effective actions, we must use the DBI and hope
that it provides an adequate description of the physics at late time.

It turns out that making the field redefinition will still not be enough to simplify the problem, and
so we are also forced to consider the throat geometry around the source branes. In terms of field space
definitions we are probing the large $T$, small $r$ region of the theory.
We can now use the Noether method to find the charge associated with a scaling symmetry on the brane world volume.
We postulate that the fields and the time scale as follows:.
\begin{equation}
t'= \Gamma^{\alpha}t, \hspace{0.5cm} r'=\Gamma^{\beta}r, \hspace{0.5cm} T'=\Gamma^{\gamma}T.
\end{equation}
Inserting these transformations into the action yields the following constraints,
\begin{equation}
\beta(p-3)= 0, \hspace{0.5cm} \alpha = -\beta, \hspace{0.5cm} \gamma = -\alpha p'.
\end{equation}
The first of these is the most important, since we have two possible solution branches. Firstly we can have
$\beta=0$, which in turn leads to $\alpha = \gamma =0$ and so there are no field symmetries. However the 
second solution gives $p=3$, which implies that the scaling variables are arbitrary.
What we have found is that the symmetry on the world-volumes of the probe branes imposes a constraint on
the allowed dimensionality of the background. If we were to allow extended transformations, for example a 
rescaling of the string coupling, we find that the background constraint becomes $p=5$. Only in the case
where we rescale all the fields, the string coupling and the string length can we eliminate this
background constraint. For simplicity, we will only look at the basic case in this note. The extension
to more general scaling symmetries is left for future endeavour.
As the scaling variables are arbitrary, we find it convenient to choose $\alpha = -1$, thus the
scalings become
\begin{equation}
t'=\Gamma^{-1}t, \hspace{0.5cm} r'=\Gamma r, \hspace{0.5cm} T'=\Gamma^{p'} T,
\end{equation}
and we find a representation of the conserved charge generating these transformations, which is 
\begin{equation}\label{concharge}
D= t \tilde E + r \tilde \Pi + p'T P_T,
\end{equation}
where $\tilde E, \tilde \Pi$ and $P_T$ are the canonical energy density, radial momentum and tachyon momentum
respectively.
Now it is useful to write the energy density in canonical form
\begin{equation}\label{hamilton}
\tilde E = \sqrt{\frac{2N^2}{T^2}\left(\frac{k_3}{r^4}\right)^{-(1+p')/2}\left \lbrace 1+
\frac{4k_3}{\lambda^2C}\right \rbrace+\frac{\tilde \Pi^2r^{4}}{k_3}+\frac{T^2P_T^2r^2}{2\lambda \sqrt{k_3}}},
\end{equation}
where we have written $k_3$ to denote the constant charge of the $D3$-brane background. Using this expression, we find the equations of
motion for the radion and tachyon fields reduce to
\begin{equation}\label{eqm}
\dot{r}=\frac{\tilde \Pi r^4}{\tilde E k_3}, \hspace{1cm} \dot{T}=\frac{T^2P_T r^2}{2\tilde E \lambda \sqrt{k_3}}.
\end{equation}
Note that in this instance, neither $\tilde \Pi$ or $P_T$ is a conserved charge which makes it difficult to
solve the equations of motion. However due to our world-sheet transformations we have discovered a charge, $D$, that is
conserved and so we can use this to simplify the equations of motion. In order to do this we will have to consider 
specific decompositions of the symmetry charge, as the general expression does not lead to simple analytic solutions.
\subsection{Decomposition of charge.}
Even with the existence of the conserved charge (\ref{concharge}) does not allow an easy split between the variables 
$r$ and $T$ which would allow us to solve the (\ref{eqm}).\footnote{The only exception is the case $p'=0 $ which we shall discuss later}. In order to try and find analytic solutions (even approximate ones) 
we need to impose further conditions on the canonical variables in a manner consistent with the equations of motion.
Let us write the conserved scaling charge $D$ in (\ref{concharge}) as the condition 
\begin{equation}
\Phi = (t \tilde{E} + r \tilde{\Pi} + p'T P_T -D) = 0
\end{equation}
This constraint is preserved under Hamiltonian flow since it can be verified that $\dot{\Phi} = d \Phi/dt +\{H,\Phi \} = 0 $
where $\{, \} $ defines the usual Poisson bracket and $H$ is the Hamiltonian defined in (\ref{hamilton}) .
Now decompose $\Phi = \Phi_1 +\Phi_2$ where
\begin{eqnarray}
\Phi_1 &= & \tilde{E}_1 t +r \tilde{\Pi} -D_1\nonumber\\
\Phi_2  &=&  \tilde{E}_2 t +p' T P_T - D_2 
\end{eqnarray}
with $ \tilde{E}_1 + \tilde{E}_2 = \tilde{E} $ and $ D_1 +D_2 = D $.
If we now impose for example, the additional constraint $\Phi_1 = 0$ (and hence $\Phi_2 = 0 $ as a consequence)
then this would allow us to solve for $r(t) $ and $T(t)$. However we must check that this additional constraint is preserved under 
Hamiltonian flow, ie that 
\begin{equation}
\dot{\Phi_1} = d \Phi_1/dt +\{H, \Phi_1 \} = 0 
\end{equation}
This leads to the following algebraic constraint between $r $ and $T$ :-
\begin{equation}\label{con2}
\tilde{E}_1 - \tilde{E} - \frac{2 N^2 p'}{\tilde{E} T^2} \left( \frac{k_3}{r^4} \right)^{-(1+p')/2} \left \lbrace 1+ \frac{4 k_3}{\lambda^2 C} \right\rbrace= 0
\end{equation}
The case $p'=0 $ is special in that the original constraint, $\Phi = 0$, can be used to solve the $r, T$ system completely 
(see later). For now we will assume that $p' \ne 0$ . Since we are considering $p' < p = 3 $  we only need consider the case when $p'=2 $.
It's clear from (\ref{con2}) that $\tilde{E}_1 > \tilde{E} $ if this constraint is to be solved exactly. But one can then show an inconsistency 
appears when this algebraic constraint is applied to (\ref{eqm}). Thus at best we can only solve (\ref{con2}) approximately. One such solution
is to take $\tilde{E}_1 \approx \tilde{E} $ and assume $T$ is large. We remind the reader that we already assumed that $T$ is large in order to 
obtain the scaling symmetries earlier. We can now go ahead and solve the $r,T$ system of equations.

Solving for the radial equation of motion we find 
\begin{equation}
\frac{1}{r^2}= \frac{1}{r_0^2}-\frac{t}{\tilde E k_3}(2D_2-\tilde E t)
\end{equation}
Now for small values of $D_2$ the dynamics of the probe obeys a $ 1/t$ relationship.
The exact description of the dynamics will depend on the relative sizes of $D_2$ and $\tilde E$. If $\tilde E >> D_2$, then
the quadratic term will be dominant. This ensures that the solution starts at some maximal distance and tends to zero. Conversely if 
$D_2$ is much larger than $\tilde E$, then the linear term is dominant and this describes an expanding solution which will break
down when the supergravity constraint is no longer satisfied. However, when the two charges are of the same order of magnitude we find a turning
solution. The sphere initially expands from $t=0$ until it reaches a stationary point at $t=D_2/\tilde E$, before collapsing toward
zero size. 

Using the second constraint to solve for the tachyon momentum yields the solution to the tachyon equation of motion
\begin{equation}
T \sim T_0 \exp\left(\frac{\sqrt{k_3}r_0^2}{4\lambda} f(t) \right)
\end{equation}
where the function $f(t)$ is proportional to $\rm{arctanh}(t\tilde E - D_2)$. Thus the general behaviour 
of the tachyon solution is that it is an exponential function of time.

The results obtained so far have all been for the case $p'=2$. In order to determine the dynamics of the $p'=0$ case
corresponding to $N$ coincident D-particles we see that the tachyon dependence drops out of of the conserved charge.
First, solving for the radial equation of motion, we find the solution
\begin{equation}
\frac{1}{r^2}=\frac{1}{r_0^2}-\frac{t}{\tilde E k_3}(2D-t\tilde E)
\end{equation}
which is a similar solution as the one obtained in the charge decomposition above for $p'=2$. Therefore
we also expect to find a similar turning solution for the fuzzy sphere parameterised by the time $t=D/\tilde E$.
We have no other constraint to impose on the equation of motion for the
tachyon field, but we can write the tachyon momentum in terms of the other canonical forms
\begin{equation}
P_T^2 = \frac{2\sqrt{k_3}\lambda}{T^2 r^2} \left(\tilde E^2 - \frac{2r^2}{T^2 \sqrt{k_3}}\left \lbrace 1 +
 \frac{4k_3}{\lambda^2C}\right \rbrace-\frac{\tilde \Pi^2 r^4}{k_3} \right).
\end{equation}
In general we can use this solution to exactly solve for the tachyon field, however this is extremely difficult 
and we will find it much more useful to find an approximate
solution. From the above equation, we see that the supergravity solution implies $r^4/k_3 << 1$, and so we can 
effectively neglect the contribution from the final two terms. Inserting this
into the equation of motion yields the solution
\begin{equation}
T \sim T_0 \exp\left( \left( \frac{\sqrt{k_3}}{2\lambda}\right)^{1/2} \rm{ln} \left[\sqrt{\tilde E k_3-
r_0^2t(2D-t\tilde E)}+\frac{r_0(tE-D)}{\sqrt{\tilde E}} \right] \right),
\end{equation}
which we expect to provide a reasonable approximation as $r\to0$, and once again shows the increasing exponential dependence of the tachyon field.
Again the contribution from the two charges can change the dynamics of the field, as described earlier.

The general solution for the tachyon field is expected to be background dependent \cite{non_bps_dynamics}, however we see that in the $D3$-case, it is
roughly exponential in all cases. The fuzzy sphere appears to always collapse, but there is an intricate relationship between the tachyon condensation and the
radial modes which depends upon the conserved charges. When both terms appear in the radial equation of motion we see that there can be turning solutions,
describing an initial expansion which eventually contracts within finite time. This is a result of the tachyon condensation which decreases the tension of the
branes so that they feel a weaker gravitational attraction. However, the combination of the charges in the tachyon solution also implies a turning point for
the tachyon field and so the tension eventually increases and the fuzzy spheres collapses - provided that the tachyon solution still remains valid.

\section{NS5-brane background.}
The work in the previous sections has only been concerned with coincident $Dp$-brane backgrounds, but we wish to
extend this to the $NS5$-brane background. This particular background is important for several reasons. 
In many cases there is an exact conformal field theory description, allowing BCFT calculations. Secondly,
there is an interesting duality which relates six dimensional string theory on the $NS$5-brane world-volume (LST) \cite{lst}
to supergravity in the bulk, permitting an understanding of the dynamics in terms of defects of the LST.
Importantly for our purposes, there has been recent work on probe dynamics in this background 
which has provided insights into the nature of the rolling tachyon, and perhaps even a geometrical origin for 
the open string tachyon in Abelian theories.
Much of the construction of the non-Abelian theory follows a similar line to that of the $D$-brane 
backgrounds.

We begin with the background solution for $k$ coincident $NS$5-branes, given by the usual CHS solutions \cite{chs}
\begin{eqnarray}
ds^2 &=& \eta_{\mu \nu} dx^{\mu} dx^{\nu} + H(x^n) dx^m dx^n \nonumber \\
e^{2(\phi-\phi_0)} &=& H(x^n) \nonumber \\
H_{mnp} &=& -\epsilon_{mnp}^q \partial_q \phi.
\end{eqnarray}
The coincident fivebranes form an infinite throat which can be seen from the dilaton term. We will
refer to the throat geometry as the near horizon part of the bulk space-time.
The usual definitions apply as in the $Dp$-brane solution, with the addition of the 3-form field
strength for the Kalb-Ramond field. The harmonic function describing this background is simply
\begin{equation}
H(x^n) = 1 + \frac{kl_s^2}{r^2},
\end{equation}
where $r$ is the physical radius given in terms of the transverse scalars, $r=\sqrt{x_i^2}$. Note that there is no WZ term in this solution, since the $NS$5-branes
are not sources of Ramond-Ramond charge. This is because the fivebrane is the magnetic dual of
the fundamental string, and as such we expect that no open strings will end on any of the
$k$ $NS$5-branes. The probe branes themselves will carry R-R charge, which we anticipate will be radiated
as the probes move in the background. This has important consequences, as we know that the classical Abelian theory
is only valid for $3 \le p < 5$ \cite{time_dependence} due to the emission of closed string modes. This tells us that the DBI only describes the 
motion of the open string degrees of freedom, and radiative correction terms due to the closed strings must be studied separately. It would be a useful
exercise to check if this relation also holds in the non-Abelian theory.
We also know that the background preserves different halves of the supersymmetry algebra, therefore it is explicitly broken
and we will find a gravitational force acting on the fuzzy sphere causing it to collapse. This is also seen in the Abelian theory
of a spherical $D2$-brane with magnetic flux \cite{hyakutake}, which should be equivalent to our construction of $D0$-branes on a fuzzy sphere.

We now insert these background solutions into our Non-Abelian action. Once again, we expand the terms to 
leading order and assume that the transverse scalars are time dependent, which will ensure
that our solutions are homogeneous and thus there will be no formation of caustics. Hence we arrive at the following form of the action
\begin{equation}
S=-\tau_{p} \int d^{p+1} \zeta STr \left( H^{-1/2} \sqrt{1-H \lambda^2 \dot{\phi}^i \dot{\phi}^j \delta_{ij}}
\sqrt{1-1/2 \lambda^2 H^2 [\phi^i, \phi^j][\phi^j,\phi^i]} \right).
\end{equation}
Note that the $NS$5-branes have a tension that goes as $1/g_s^2$, whilst the $Dp$-branes each have tensions proportional to $1/g_s$, thus
the five-branes are heavier in the large $k$ limit, however as we will be interested in the large $N$ limit we may find there is considerable back reaction
upon the throat in the target space which may deform it substantially. However for the purpose of this note we will ignore this effect, and simply assume
that we can fine tune the parameters such that the back reaction is negligible. 
The action is given by \footnote{For simplicity we do not include angular momentum though this can be done
as in section 4}

\begin{equation}
S=-\tau_p \int d^{p+1} \zeta \frac{N}{\sqrt{H}}\sqrt{(1-H\lambda^2 \dot{R}^2C)(1+4\lambda^2 CH^2 R^4)},
\end{equation}

with $C$ being the usual quadratic Casimir of the $N$-dimensional representation. Switching now to physical
distances, we arrive at the final form of the action
\begin{equation}
S=-\tau_p \int d^{p+1} \zeta \frac{N}{\sqrt{H}} \sqrt{\left(1-H \dot{r}^2\right)\left(1+\frac{4H^2r^4}{\lambda^2C}\right)},
\end{equation}
from which we can derive the usual canonical momenta and energy densities, where we have explicitly divided out
the 'mass' of each brane.
\begin{eqnarray}
\tilde \Pi &=& \frac{NH\dot{r}}{\sqrt{H}}\sqrt{1+\frac{4H^2r^4}{\lambda^2C}} \frac{1}{\sqrt{1-H \dot{r}^2}} \nonumber \\
\tilde E &=& \frac{N}{\sqrt{H}}\sqrt{1+\frac{4H^2r^4}{\lambda^2C}} \frac{1}{\sqrt{1-H \dot{r}^2 }}.
\end{eqnarray}
We solve the equation for the energy, which is conserved, to obtain the following
constraint on the dynamics of the probe branes assuming a  fixed energy density
\begin{equation}
1 \ge \frac{N^2}{ \tilde E^2 H} \left(1+\frac{4H^2r^4}{\lambda^2C}\right)
\end{equation}
We are going to be interested in the near horizon geometry of the fivebrane background, and so
can make the usual approximation with regards to the harmonic function. Again we will also anticipate that
there is a maximum size for the fuzzy sphere in the large $r$ region, since in this region the metric reduces
to the metric for the $Dp$-brane background, namely Minkowski space.
In the throat, we find that the constraint becomes 
\begin{equation}
1 \ge \frac{N^2 r^2}{\tilde E^2 k l_s^2} \left(1+\frac{4k^2l_s^4}{\lambda^2C}\right)
\end{equation}
which is automatically satisfied for the radial part since we know that $H >> 1$ in this region.
This actually allows us to find the following constraint on the energy density
the following constraint on the energy
density
\begin{equation}
\frac{\tilde E^2}{N^2} \ge \frac{r^2}{kl_s^2}\left(1+\frac{k^2}{\pi^2C}\right).
\end{equation}
The supergravity solution tells us that $r^2/kl_s^2$ must be extremely small, and we can select $k/N$ to be 
small even when $k$ and $N$ are individually large, thus the last term is simply $\mathcal O(1)$
This implies that we should take $\tilde E$ to be larger than $N$ to ensure that the constraint is satisfied.
Thus like the majority of the $Dp$-brane solutions we find that
the fuzzy sphere can collapse down toward zero size. In this background though we expect that the moving $D$-branes will shed their energy, which
will appear as modes living on the fivebranes, and eventually form a $(k, N)$  bound state in analogy to the $(k,1)$ state in the Abelian case.
As we have already seen, one of the main differences between the usual fuzzy sphere solutions in flat space and those in curved background is that the 
velocity term decreases with the radius. In flat space we find that the collapsing configuration approaches the speed
of light at late times and thus the $1/N$ corrections due to the symmetrized trace become important. Clearly we don't see the
same behaviour in this case, in fact a six dimensional observer on the
$NS$5-brane world volume will record that it takes an infinite amount of time for the fuzzy sphere to collapse to zero size
\footnote{Of course if we switch to 'proper' co-moving time coordinates $\tau$, then the collapse will occur in finite time \cite{time_dependence}, and an observer
on the probe branes will record the velocity as tending to the speed of light.}
. This is interesting as it appears that the energy of a collapsing fuzzy sphere in flat space is the same as an essentially static
sphere in a space-time throat, and is related to the formation of a bound state of $(p,q)$ fivebranes \cite{time_dependence}.
In the large $r$ region we find
\begin{equation}
1 \ge \frac{4N^2r^4}{\tilde E \lambda^2 C},
\end{equation}
which translates into the condition that the fuzzy sphere has a maximum radius given by
\begin{equation}
r_{max} = \sqrt{ \frac{ \tilde E \lambda C^{1/2}}{2N}}.
\end{equation}
Which, as anticipated, is the same result derived for the $D$-brane background.
We now look at the static potential associated with the fivebrane background. Following the convention employed in the Abelian cases \cite{time_dependence}, we easily find
that the potential can be written
\begin{equation}
V_{eff} = \frac{N^2}{\tilde E^2 H^2}  \left(1+\frac{4H^2r^4}{\lambda^2C}\right) - \frac{1}{H}
\end{equation}
The interesting question is what happens in the throat, since we know that in the large $r$ region
the potential will be a simple monotonically increasing function, which goes as $r^4$
\begin{equation}
V_{eff} \sim \frac{4r^4}{\lambda^2 \tilde E^2}.
\end{equation}
 Dropping the factor of unity as before, we
find that as $r \to 0$ the potential becomes
\begin{equation}
V_{eff} \approx \frac{N^2 r^4}{\tilde E^2 k^2 l_s^4} \left(1+\frac{4k^2l_s^4}{\lambda^2 C}\right) - \frac{r^2}{kl_s^2}.
\end{equation}
which indeed tends to zero with $r$, for fixed $\tilde E$ as expected

In any case, We wish to solve the equation of motion for the probe branes in the throat. Because the energy is conserved, the solution, up to constants of integration, is simply
\begin{equation}
\frac{1}{r} = \sqrt{ \frac{N^2}{\tilde E^2 kl_S^2}\left(1+\frac{4k^2l_s^4}{\lambda^2c}\right)}\cosh\left(\frac{t}{\sqrt{kl_s^2}}\right),
\end{equation}
which is actually an extension of the solution for a single probe $Dp$-brane in the Abelian theory, which was shown to be \cite{time_dependence}
\begin{equation}
\frac{1}{r}=\frac{1}{\sqrt{\tilde E^2 kl_s^2 - \tilde L^2}}\cosh\left(\frac{t}{\sqrt{kl_s^2}}\sqrt{1-\frac{\tilde L^2}{kl_s^2\tilde E^2}} \right).
\end{equation}
where the effect of the angular momentum is to act as a 'braking' term. 

If we consider performing a Wick rotation of the time coordinate for the collapsing
solution we find a periodic solution in terms of a cosine function. This can be interpreted as the collapse
and subsequent bounce of the fuzzy sphere in imaginary time - although the physical interpretation of this solution is not clear, however we
expect it to approximate the time dependent solution for Euclidean branes.
This sinusoidal behaviour can also be seen if we switch to conformal (or 'proper') time where an observer sees that the collapse occurs in
finite time. In this case we would expect $1/r$ to be proportional to $\sin(t)$ \cite{time_dependence} which again is suggestive of a periodic
collapse and expansion. However this solution would indeed probe the non-perturbative region of the theory, and it is not clear if the corrections
(e.g quantum, 1/$k$ and back-reaction) would admit such a solution. One further thing to note is that using S-duality we may map this solution to
that of the coincident $D5$-brane background being probed by coincident $D3$-branes, as their actions are identical. This agrees with our expectation
that the $D5$-brane background yields exponential solutions at late times.

We may enquire about the validity of the classical solution in the throat region. Using our time dependent ansatz we see that the dilaton is also a time dependent function, in fact for a purely
radially collapsing solution we find that the dilaton behaves as
\begin{equation}
e^{\phi} = \frac{Ng_s}{\tilde E}\sqrt{1+\frac{4k^2l_s^4}{\lambda^2 C}}\cosh \left(\frac{t}{\sqrt{kl_s^2}} \right).
\end{equation}
Note that quantum effects can be neglected provided that $e^{\phi} << 1$, however as we know that $\tilde E>> N$ from
our constraint equation we expect that the classical analysis will provide an accurate description of the solution, at least for
early times. This can be 'fine tuned' for specific values of $k$ and $N$ so that the classical solution continues to hold at late
times.
\subsection{Correction from symmetrized trace.}
Thus far we have investigated the dynamics of the action at leading order, and seen that
the fuzzy sphere will generally collapse down to small size. It is expected that the effective action
will break down at distances comparable with the string length, and thus $1/N$ corrections will become
important. In order to deal with this situation we look at the next order terms due to the
symmetrized trace corrections.
As we have already seen, we can write the first order correction to the energy as
\begin{equation}
\tilde E_1 = \left(1-\frac{2}{3}C\frac{\partial^2}{\partial C^2}\right) \tilde E_0,
\end{equation}
which yields
\begin{equation}
\tilde E_1 = \frac{N}{\sqrt{H}\sqrt{1-H\dot{r}^2}} \left(W(k,C) + \frac{2Ck^4}{3W(k,C)^{3/2}\pi^4 C^4} - \frac{2k^2C}{3W(k,C) \pi^2 C^3}  \right).
\end{equation}
Where, for simplicity, we have re-introduced the notation
\begin{equation}
W(k,C) = \sqrt{1+\frac{k^2}{\pi^2 C}}.
\end{equation}
This term can be thought of as a mass term, by seeing how it arises in the context of the energy. In the $Dp$-brane case
(and in flat space) this term will be position dependent, and we have the notion of a position dependent mass. However,
the near horizon of the $NS$5-brane background removes the radial dependence leaving us with a constant. Because we are using
the supergravity approximations in our analysis, we are taking $k$ and $N$ to be large, and so this 'mass' term is positive,
but may be small if we demand that the ratio $k/N$ be small.
If we now employ the canonical formulation of the energy, we can set the $\tilde \Pi$ terms to zero to find the 
corrected potential for the probe branes up to leading order in $1/C$
\begin{equation}
V_1 = \frac{N}{\sqrt{H}}\left( W(k,C) -\frac{2k^2}{3W(k,C) \pi^2 C^2} \right).
\end{equation}
The potential does not vanish with this correction because we have the supergravity condition $k>>N$ where both
$k$ and $N$ are integers. In fact, even taking into account higher order corrections \cite{ramgoolam}, the potential
is nowhere vanishing. Thus the symmetrized trace correction does not yield a bounce solution.
\subsection{Non-Abelian tachyon map.}
It has been shown in the case of a single probe brane, that the unstable dynamics in the $NS$5-brane background is more easily understood in 
terms of the rolling tachyon, since the energy momentum tensors have similar behaviour at late times.
We may ask what the implications are when we have multiple coincident branes with a $U(N)$ symmetry on their worldvolumes.
This relationship can be
explicitly demonstrated by mapping the probe brane action into that of the tachyon action in flat Minkowski space. 
This is particularly simple in the Abelian case, but we wish to show that it is also possible in our
non-Abelian construction.
The corresponding non-Abelian action for tachyons in a flat background, to leading order \cite{non_abelian_non_bps}, can be written
\begin{equation}
S=-\tau_p V_p \int dt N V(T)\sqrt{1-\dot{T}^2}.
\end{equation}
Because the tachyon field does not take values in the $SU(2)$ algebra we find that the action is simply $N$ times that
of a single non-BPS brane. In fact this corresponds to a configuration of branes each separated by distances
larger than the string length, as found in constructions of Assisted Inflation \cite{assisted_inflation}. In this scenario each of the tachyons is
assumed to follow a similar trajectory toward the late time attractor point, namely $T_1 \sim T_2 \ldots \sim T_N \equiv T$.
Here $V_p$ is the effective 'volume' of each brane, whilst $V(T)$ is the tachyon potential which we will assume to be of the form;
\begin{equation}
V(T)=\frac{1}{\cosh(T/T_0)},
\end{equation}
where the tachyon is a scalar field with dimensions of length. We remind the reader of the action for the probe brane in 
the $NS$5-background, which we have already show to be
\begin{equation}
S=-\tau_p V_p \int dt \frac{N W(r)}{\sqrt{H}} \sqrt{1-H \dot{r}^2},
\end{equation}
where, for simplicity, we have absorbed the potential term into our definition
of $W(r)$.
Clearly we can map this action to that of the non-Abelian tachyon by making the identification
\begin{equation}
d\tilde T = \sqrt{H} dr, \hspace{0.5cm} V(\tilde T) = \frac{W(r)}{\sqrt{H}}.
\end{equation}
Using the near horizon approximation we can solve for the geometrical tachyon in terms of the physical radius
of the fuzzy sphere. The result, up to arbitrary constants of integration, is simply
an exponential as expected from the Abelian case which allows us to write the tachyon field as
\begin{equation}
\tilde T \sim \sqrt{kl_s^2}\ln(r).
\end{equation}
The solution tells us that as $r \to 0, \tilde T \to -\infty$ as expected, whilst
as $r \to r_{max}$ we find $\tilde T \to \tilde T_{max}$. Clearly this is not
the general behaviour associated with the open string tachyon solution, which we should
have expected from the Abelian theory,
but we may anticipate that the decay of the fuzzy sphere will also be 
describable in terms of this rolling tachyon solution.
Using our field redefinition, we write the expression for the tachyon potential as
\begin{equation}
V(\tilde T) = \sqrt{\frac{1}{kl_s^2} \left( 1 + \frac{k^2}{\pi^2 C} \right)}
\exp \left( \frac{\tilde T}{\sqrt{kl_s^2}}\right).
\end{equation}
The form of the potential shows that it had its maximum at $\tilde T=0$, and tends to zero for $\tilde T \to -\infty$. 
The exact maximum will be defined by the
number of source branes, as expected from the Abelian case. However note that there is a correction
term present here due to the fuzzy sphere, which does not occur in the leading order tachyon action
as we know that the tachyonic scalar field is a commuting variable. Therefore although we can capture the general
behaviour of the tachyon action, we must go beyond leading order to find closer agreement.
If we construct the Energy-Momentum tensor associated with this rolling tachyon solution, omitting the
delta functions which localise the tensor on the brane world-volumes, we obtain
\begin{eqnarray}
T_{00} &=& \frac{NV(\tilde T)}{\sqrt{1-(\partial_t \tilde T)^2}} \nonumber \\
T_{ij} &=& -NV(\tilde T) \sqrt{1-(\partial_t \tilde T)^2},
\end{eqnarray}
which shows that the pressure tends to zero as the potential tends to zero, ie,
when the probe branes approach the fivebranes at late times. This is because the probe brane will emit 
energy in closed string modes as the fuzzy sphere collapses,
and the resulting matter will be Non-Abelian pressureless fluid. One must also
imagine that because the fuzzy sphere collapses in the near throat region of the
fivebranes, becoming pointlike at distances
approaching the string length, the harmonic function approximation may fail, and there will certainly be
quantum corrections to take into account. 
This is due in part to the back reaction of the probes on the source branes and the throat,  therefore in order to
determine the physics of this non-Abelian fluid it will
be necessary to calculate this back reaction term and incorporate it into the action. In any case, it would be useful
to compute the dynamics of this configuration using the exact CFT on the world volume which would help shed further light on
the validity of the classical solution.
\section{Non-BPS branes in fivebrane backgrounds.}
We now wish to extend our discussion to include non-BPS branes in this
coincident fivebrane background. As is well known the BPS $Dp$-brane is a soliton
solution of the non-BPS $D(p+1)$-brane, where the soliton is associated with
the condensation of the open string tachyon on the world-volume.
We begin by introducing the natural
extension of the Abelian action for $N$ Non BPS branes \cite{non_abelian_non_bps}.
\begin{equation}
S=-\int d^{p+1} \zeta Str  V(T) e^{-(\phi-\phi_0)}
\sqrt{-\det (\mathcal{P}[ E_{ab}+E_{ai}(Q^{-1}-\delta)^{ij}E_{jb}]+\lambda
F_{ab}+T_{ab})} \nonumber
\end{equation}
\begin{equation}
\times \sqrt{\det Q_j^i}
\end{equation}
where $T_{ab}$ is the tensor containing all the open string tachyon terms, and is given by
\begin{equation}
T_{ab}=\lambda D_a T D_b T -D_a T[x^i,T](Q^{-1})_{ij}[x^j,T]D_b T+ \ldots.
\end{equation}
Note that we are now taking the tachyon to be a dimensionless scalar field on the
world-volumes of the $N Dp$-branes by reinstating the factors of $\alpha'$.
We now expand the action to lowest order, and we will drop the gauge field
term so that covariant derivatives reduce to normal derivatives.
The resulting action can be written
\begin{equation}
S=-\int d^{p+1}\zeta STr \frac{V(T)}{\sqrt{H}}\sqrt{\left(1+H\lambda^2
\partial_0 \phi^i \partial^0 \phi^j \delta_{ij}+\lambda \partial_0 T
\partial^0 T\right) \left(1-\frac{1}{2} H^2 [\phi^i, \phi^j][\phi^j,\phi^i]\right)}.
\end{equation}
We again use the $SU(2)$ ansatz for the radially dependent transverse scalars which reduces
the action to a more tractable form
\begin{equation}
S=-\int d^{p+1}\zeta N\frac{V(T)}{\sqrt{H}}\sqrt{(1-H\lambda^2C \dot{R}^2 -\lambda \dot{T}^2) 
(1+4H^2\lambda^2 C R^4)}.
\end{equation}
The presence of the open string tachyon will generally prohibit exact solutions
to the equations of motion for the radion field unless we take various asymptotic limits.
This is obvious, as the form of the action shows that the conjugate momenta
associated with the radion and tachyon fields will not be conserved.
Fortunately there is a way to resolve this problem by using symmetry transformations of the 
various fields, which allows us to construct a new conserved charge and hence
solve the equations of motion for specific regions of field space.

We will start by considering the usual form of the tachyon potential for the superstring, given by
\begin{equation}
V(T)=\frac{1}{\cosh(T/ \sqrt{2})},
\end{equation}
which tends to an exponential for large $T$ in agreement with calculations from string field theory and BCFT \cite{sen}.
We insert this into the action, and once again switch to using physical distance. We note that
the current form of the potential will make it difficult to find symmetries of the action as it
stands, thus it will be more useful to make the following field redefinition \cite{tachyon_transform, non_bps_dynamics} as we did for the coincident $D$-brane
background
\begin{equation}
\frac{\tilde T}{\sqrt{2}} = \sinh \left(\frac{T}{\sqrt{2}}\right),
\end{equation}
and for convenience we re-write $\tilde T = T$ for ease of calculation, although we will always
imply that this is the re-definition of the original tachyon field. As mentioned previously there may be objections to performing 
this kind of field redefinition using the non-Abelian action in this gravitational background.
Assuming that this won't be too problematic, we can now proceed to analyse the resulting action,
\begin{equation}
S=-\tau_p \int d^{p+1} \zeta \frac{N}{\sqrt{HF}} 
\sqrt{\left(1-H\dot{r}^2-\frac{\lambda \dot{T}^2}{F}\right)\left(1+\frac{4H^2r^4}{\lambda^2C}\right)},
\end{equation}
where we have introduced the following definitions
\begin{equation}
F=\left(1+\frac{T^2}{2} \right), \hspace{1cm} H=1+\frac{kl_s^2}{r^2}.
\end{equation}
We can now try to find the conserved charge associated with transformations of this action, and use that in conjunction with the
energy density to solve the equations of motion.
Unfortunately we see that this is still non trivial unless we make further
approximations, thus we will look at the theory in the large $T$ and small $r$ limit. Since the large tachyon field
gives rise to a gas of closed strings arising due to tachyon condensation, we expect to find
that the radial field on the probe branes will describe the late time dynamics of this gas.
The action in this instance, reduces to
\begin{equation}
S=-\tau_p \int d^{p+1} \zeta \frac{\sqrt{2}Nr}{\sqrt{kl_s^2}T}\sqrt{1-\frac{kl_s^2 \dot{r}^2}{r^2}-\frac{2\lambda\dot{T}^2}{T^2}}
\sqrt{1+\frac{k^2}{\pi^2C}},
\end{equation}
At this juncture we will reintroduce the $W(k,C)$ notation to simplify things, and furthermore, 
we postulate that the action be invariant under the following transformations
\begin{equation}
\delta T = \epsilon T, \hspace{1cm} \delta r= \epsilon r,
\end{equation}
for some parameter $\epsilon$. Note that this is a transformation involving the open strings on the world-volume and
also the transverse scalars, and can be thought of as an example of the stringy space-time uncertainty principle \cite{uncertainty_principle}
\begin{equation}
\Delta t \Delta X \ge \alpha'
\end{equation}
where distances on the world-sheet are inversely related to distances in the bulk. Since the $NS$5-brane
world-volume theory is related to a Little String Theory (LST), it would be interesting to find out
the implications of the  transformations on the LST side.
By variation of the action, we determine that the charge associated with this symmetry is given by
\begin{equation}
D = \frac{ N r \sqrt{2}}{T\sqrt{kl_s^2}} \left(\frac{kl_s^2 \dot{r}}{r}+\frac{2\lambda \dot{T}}{T}\right)
\frac{W(k,C)}{\sqrt{1-\frac{kl_s^2 \dot{r}^2}{r^2}-\frac{2\lambda \dot{T}^2}{T^2}}},
\end{equation}
which can be seen to have dimensions of length.
We can also derive the canonical energy density associated with the action, using the canonical momenta
of the radion and the tachyon fields. For brevity we will simply state the resultant dimensionless energy density and
not the individual momenta
\begin{equation}
\tilde E= \frac{Nr\sqrt{2}W(k,C)}{T\sqrt{kl_s^2}} \frac{1}{\sqrt{1-\frac{kl_s^2 \dot{r}^2}{r^2}-\frac{2\lambda \dot{T}^2}{T^2}}}.
\end{equation}
It can be seen that both $\tilde E$ and $D$ are conserved, as expected, and it will be useful to combine both of these
charges to form a solitary conserved charge
\begin{equation}\label{eq:charge}
Q=\frac{D}{\tilde E}=\frac{kl_s^2 \dot{r}}{r} + \frac{2 \lambda \dot{T}}{T},
\end{equation}
which after some manipulation can be used to define the tachyon field via
\begin{equation}
T= C_o \exp\left(\frac{Qt}{2\lambda}\right)r^{-k/4\pi},
\end{equation}
where $C_0$ is a constant of integration. Furthermore, from (\ref{eq:charge}) we can also find
the time dependence of the tachyon field in this condensing limit.
\begin{equation}
\dot{T} = T \left( \frac{Q}{4\pi l_s^2}-\frac{k\dot{r}}{4\pi r} \right).
\end{equation}
As we are probing the large $T$ region of field space, we expect that the dominant contribution to the charge
will come from the radial modes.
Now that we have written the tachyon field in terms of this conserved charge we can attempt to solve
the radial equations of motion. Note that this would be extremely challenging if we had tried to proceed
from the original form of the action without this enhanced symmetry.
We will initially consider the case where $Q=0$. This obviously implies that we are setting $D \to 0$, which
may seem strange, however we have used the charge to construct an expression for the tachyon field and 
so it is valid. By setting $Q=0$ we are identifying the condensation of
the tachyon field with the inverse of the radion field on the probe branes (up to some power), and so small $r$ will automatically imply large $T$.
The simplicity of this approach is now clear, since we began with two distinct fields and have effectively coupled them via the conserved charge thus
only requiring us now to solve for one of the fields.
We now substitute the expressions for the tachyon into the energy equation, which will
now be a function of $r$. 
\begin{equation}\label{eq:energy_solution}
\tilde E=\frac{NW(k,C)r^y\sqrt{2}}{C_0\sqrt{kl_s^2}}\frac{1}{\sqrt{1-\frac{kl_s^2\dot{r}^2}{r^2}\left(1+\frac{k}{4\pi}\right)}}
\end{equation}
and for future reference, we will introduce the simplifying notation
\begin{equation}
B= \frac{N W(k,C) \sqrt{2}}{\tilde E C_0 \sqrt{kl_s^2}}, \hspace{0.5cm} y= 1+\frac{k}{4\pi}, \hspace{0.5cm} x=kl_s^2\left(1+\frac{k}{4\pi}\right)
\end{equation}
Rearranging the energy  equation allows us to solve for $r(t)$, which we find to be, up to constants of integration
\begin{equation}
\frac{1}{r}  \sim \left(B \cosh\left\lbrack\frac{\pm y(t-t_0)}{\sqrt{x}}\right\rbrack \right)^{1/y},
\end{equation}
where $t_0$ parameterises some initial time for the dynamics.
This solution describes an expanding fuzzy sphere which reaches its maximum size at $t=t_0$ and thereafter collapses down to zero size.
 We easily find that the maximum radius will be given by
\begin{equation}
r_{\rm max} = \left(\frac{\tilde E C_0 \sqrt{kl_s^2}}{NW(k,C)\sqrt{2}}\right)^{1/y}.
\end{equation}
The physics behind this solution can be understood.  As the fuzzy sphere expands, the tension of the 
non-BPS branes is increased as the tachyon moves closer to the top of its potential (assumed to be located at $T=0$ ).
Thus the expanding solution has a natural braking force that restricts it to expand to a certain size.
Conversely in the collapsing phase, the non-BPS branes feel a decreasing tension which goes to zero as the solution collapses to the origin. 

We can also now determine the constant of integration by demanding that $T=T_0$ at $t=t_0$, and since
we are in the large $T$ region of field space we will assume that $|T_0| >> 1$. After some 
manipulation we find
\begin{equation}
C_0 = T_0^{y}\left(\frac{NW(k,C) \sqrt{2}}{\tilde E l_s^2 \sqrt{k}}\right)^{k/ 4\pi  },
\end{equation} 
and therefore we can completely determine the behaviour of the tachyon near condensation, in the approximation
where $Q=0$. It is natural to now consider the case where $Q \ne 0$, however we should note that this
case is not solvable exactly, and we must be forced into making approximations,r
If we insert the full expression for the tachyon field into the energy equation we find
\begin{equation}
1-\frac{kl_s^2\dot{r}^2}{r^2}-\frac{l_s^2}{4\pi}\left( \frac{Q^2}{l_s^4}-\frac{2Qk\dot{r}}{l_s^2r}+\frac{k^2\dot{r}^2}{r^2} \right)
=B^2 e^{-Qt/\lambda} r^{2y}.
\end{equation}
Now at late times we see that the RHS of this equation will become vanishingly small, and so we neglect it in
our analysis. This allows us to rewrite the LHS as a quadratic equation, which we solve to find
\begin{equation}\label{eq:beta}
\frac{\dot{r}}{r}= \frac{Qk \pm 2\sqrt{k\pi (4\pi l_s^2+kl_s^2-Q^2)}}{(4\pi k+k^2)l_s^2}  =\beta,
\end{equation}
and upon integration we can determine the late time behaviour of the fuzzy sphere
\begin{equation}\label{eq:gen_rad_soln}
r \simeq r_0 \exp(\beta t),
\end{equation}
with the corresponding late time solution for the tachyon field given by
\begin{equation}
T \simeq \exp\left(\frac{Qt}{4\lambda}\right)\exp(-k\beta t /8\pi ).
\end{equation}
Now if we look for a collapsing solution we must take $\beta$ to be negative in (\ref{eq:gen_rad_soln}), where we must bear
in mind that the solution is only valid for large $t$, corresponding to late time dynamics of the radial modes.
In this case the tachyon field will be large even if the charge $Q$ is small, and so our analysis is consistent.
Furthermore having non-zero $Q$ appears to imply that there will not be a bounce solution, rather the probe branes
will eventually reach the source branes and the fuzzy sphere will collapse to zero size. 
This can be seen from (\ref{eq:beta}) which suggests that for a real
solution, we must ensure that $(4 \pi +k)l_s^2 \ge Q^2$. In the large
$k$ limit this is approximated by the constraint $kl_s^2 \ge Q^2$. Clearly if this is saturated then we find
\begin{equation}
\beta \to \frac{Q}{(4\pi +k)l_s^2},
\end{equation}
which is dependent upon the sign of $Q$. If we accept the constraint, then for $\beta$ to be negative
we require 
\begin{equation}
4\pi (4\pi l_s^2 + kl_s^2-Q^2) > Q^2k ,
\end{equation}
which becomes
\begin{equation}
4\pi l_s^2 > Q^2
\end{equation}
when we ensure $k>>1$.
Clearly the only way to satisfy this constraint is to assume that $Q$ is vanishingly small. 
This is inconsistent with (\ref{eq:charge}) for both expanding and contracting solutions.

One way of interpreting the physical aspect of the conserved charge is that it parameterises the deviation from the
single field duality we found when we identified the tachyon field with the inverse of the radial mode.

\section{Higher (even) dimensional fuzzy spheres.}
So far our analysis has dealt with collapsing fuzzy $2$-spheres in curved backgrounds, thus it would be useful to extend this
to higher dimensional fuzzy spheres. We will briefly look at the fuzzy $4$-sphere before commenting on how our analysis can
generalise to the fuzzy $2k$ sphere where $k$ is an integer. In the following discussion we will concern ourself with $D$-brane
backgrounds for simplicity, as to consider the $NS$5-brane background we will have to use T-duality.

We begin by constructing the fuzzy $\mathbb{S}^4$, where we need five transverse scalar fields satisfying the following
ansatz
\begin{equation}
\phi^i = \pm R G^i, \hspace{1cm} i=1 \ldots 5.
\end{equation}
This will obviously imply that we can only look at $p\le 4$ backgrounds. The $G^i$ matrices above arise through the totally symmetric $n$-fold tensor product 
of the gamma matrices of $SO(5)$, which have dimension
\begin{equation}
N =\frac{(n+1)(n+2)(n+3)}{6}.
\end{equation}
For a detailed description of these constructions we refer the interested reader to \cite{myers2, ramgoolam2, costis} and the references therein
In terms of the physical radius we find a similar relationship to the case of the $SU(2)$ algebra, where we write
\begin{equation}
r = \lambda \sqrt{C} R,
\end{equation}
note that in this instance $R$ must be positive definite and the Casimir is given by products of the $G^i$, as usual,
where we have $G^i G^i = C\mathbf{1}_{N \times N} = n(n+4) \mathbf{1}_{N \times N}$. We can now use this ansatz in our
non-Abelian DBI effective action, which we again treat as a lowest order expansion. The resultant action may be written
\begin{equation}
S = -\tau_{p'} \int d^{p'+1} \zeta N H^{(p-p'-4)/4} \sqrt{1-H\lambda^2 C \dot{R}^2}\left(1+4H\lambda^2CR^4 \right) - \tau_{p'} \delta_{p p'} \int d^{p'+1} \zeta \frac{qN}{H},
\end{equation}
where the Chern-Simons term only couples to the action for $p=p'$ as usual. From this action we can derive the usual canonical momenta and 
energy, which yields the following static potential in terms of physical distances
\begin{equation}
V_{eff} = \tau_{p'} N H^{(p-p'-4)/4} \left(1+\frac{4Hr^4}{\lambda^2 C} \right),
\end{equation}
note that this appears to gave exactly the same basic structure as the fuzzy $\mathbb{S}^2$ potential except that now $p \le 4$ because
of our ansatz.
Before we comment on this solution, we should discuss the extension to the fuzzy $\mathbb{S}^6$. We again use the $G^i$ matrices
which are now representations of $SO(7)$ as $i$ runs over seven transverse dimensions. Again the $G$'s arise from the action of gamma
matrices on the traceless, symmetric $n$-fold tensor product of the spinor, and we have the following relationship between the dimension
of the matrices and the number of branes
\begin{equation}
N = \frac{(n+1)(n+2)(n+3)^2(n+4)(n+5)}{360}.
\end{equation}
The relationship between the physical radius and the transverse scalar ansatz is the same as before except that the Casimir has a different
definition $G^i G^i = C\mathbf{1}_{N \times N} = n(n+6) \mathbf{1}_{N \times N}$. This suggests that we can make the following generalisation.
For the fuzzy $\mathbb{S}^{2k}$ sphere  in ten dimensions, where $k \le 4$ we require $2k+1$ transverse scalar fields which can be parameterised by
the action of $SO(2k+1)$ gamma matrices on tensor products of the spinor. If we assume that this is correct then we propose to write the general
form of the action for fuzzy $\mathbb{S}^{2k}$ in a curved $D$-brane background \cite{costis}
\begin{equation}
S = -\tau_{p'} \int d^{p'+1} \zeta N H^{(p-p'-4)/4} \sqrt{(1-H \lambda^2 C_k \dot{R}^2)(1+4H\lambda^2 C_k R^4)^k} - \tau_{p'} \delta_{p p'} \int d^{p'+1} \zeta \frac{qN}{H}.
\end{equation}
Where we have written $C_k$ to indicate that the Casimir refers to the gauge group $SO(2k+1)$.
The factor of $k$ imposes restrictions upon the dimensionality of the background branes, in fact the maximum value of $p$ is $p_{max}=8-2k$. Thus we 
see that for the fuzzy $\mathbb{S}^8$ we can only consider $D0$-branes probing the $D0$-brane background.
Using the general form of the action we define the effective potential, in physical coordinates, to be
\begin{equation}
V_{eff} = N \tau_{p'} \left\lbrace H^{(p-p'-4)4}\left(1+\frac{4Hr^4}{\lambda^2 C_k} \right)^{k/2} -q \delta_{p p'}  \right\rbrace.
\end{equation}
In general we see that the bosonic part of the potential will always tend to zero in the near horizon region, implying that the fuzzy spheres will collapse
toward zero size. Thus the only case of interest relates to $p=p'$ when there is the additional term coming from the bulk RR charge of the background branes. In the small radius limit we 
find that the potential reduces to
\begin{equation}
V_{eff} = \frac{N \tau_{p'}}{H} \left\lbrace \left(1+ \frac{4k_p r^{p-3}}{\lambda^2 C_k} \right)^{k/2} -q \right\rbrace.
\end{equation}
We can differentiate this potential to see if there are any solutions corresponding to stable mimima at which point the 
fuzzy sphere may stabilise, however we see that there are no real solutions again implying that all fuzzy spheres
are unstable in $D$-brane backgrounds with the exception of $p=6, p'=0$ which we discussed in a previous section.

The generalised form of the equation of motion can be written as
\begin{equation}
\dot{r}^2 = \frac{1}{H} \left\lbrace 1-\frac{N^2 \tau_{p'}^2 H^{(p-p'-4)/2}}{E^2} \left(1+\frac{4Hr^4}{\lambda^2 C_k} \right)^{k/2} \right\rbrace,
\end{equation}
where we are using a generalised expression for the energy. If we again assume that the velocity and the radius can be treated as complex variables with the
equation of motion  as a constraint, we can calculate the genus of the underlying Riemannian surface. Interestingly the results similar to those obtained
in section 3, with the number of branch points being the same, though the the genus is dependent upon the dimensionality of the branes and on the non-Abelian group structure.
This may change once corrections to the symmetrized trace are taken into account.
\section{Discussion.}
In this note we have extended the work on time dependent solutions to include multiple probe branes via the 
non-Abelian effective DBI action. In particular we have focused on the dynamics of BPS and non-BPS branes in the
curved backgrounds of $Dp$-branes and $NS$5-branes. This preliminary analysis has not dealt with the difficult problem of
$Dp'$-branes in the $Dp$-background, nor the fundamental string background - which is exactly soluble in the Abelian
case. It would certainly be useful to continue this line of enquiry in the future. It would also be useful to consider
ring backgrounds, both for the $Dp$-brane and $NS$5-brane backgrounds as a natural extension of the work in
\cite{israel, asano, huang, time_dependence} which could shed further light on the geometrical nature of the tachyon. The $D6$ ring could
also be an interesting case to study, as we could imagine a more general construction of a toroidal QHS.

We have seen that the fuzzy sphere is not generally a stable object when placed in non-trivial backgrounds 
(the exception is the $D6-D0$ system). Of course, this has only 
been investigated to leading order and there are many ways in which to stabilise the spheres using fluxes \cite{myers}. Furthermore,
we have only treated the configuration as a probe of the geometry. In more realistic scenarios there will be considerable
back reaction upon the background which needs to be taken into account, as well as quantum corrections when the
radius becomes significantly small. We also know that the classical motion of $D$-branes in $NS$5-brane background \cite{time_dependence} 
has a potential divergence in the closed string energy emission. We have not calculated this term here, but it would
be interesting to verify that the same thing occurs in the non-Abelian picture, and also determine whether this imposes any
constraints upon the dimensionality of the probe branes. 
Additionally we have looked at the underlying geometry of the holomorphic differentials in curved backgrounds, which suggests that they correspond to surfaces
of high genus which are clustered near the origin and thus unresolvable when we are far away, mimicking the results obtained
in flat space \cite{costis}. The genus of the particular surface is dependent upon the dimensionality of the branes in the bulk
spacetime. Smaller values of $p$ correspond to surfaces of higher genus, however larger values of $p-p'$ correspond to surfaces of smaller
genus. In the Abelian theory we see that there appears to be a relationship between the existence of supersymmetry and a Riemmanian sphere,
which is broken when we lift to the non-Abelian theory. The underlying reasons for this are unclear as the $p=6, p'=2$ solution now describes 
a torus as opposed to the sphere.
Furthermore we know that the automorphisms of the curves in flat space are destroyed when we 
move into the near horizon geometry, and the large-small dualities between collapsing and static solutions must be modified
accordingly. We would like to know whether this duality holds (albeit modified) in the throat geometry, and what are the implications
for the automorphisms. 

By careful choice of ansatz for the non-commuting coordinates, we have been able to study a rotating fuzzy sphere. In the first instance we
were able to find an expansion of the action allowing for small angular momentum densities, but only for $p \le 5$. This showed that there
were no bound states permitted for the fuzzy sphere. A second ansatz for general angular momentum imposed stricter restrictions upon
the dimensions of the branes, limiting us to $p \le 3$. Again, we saw no possibility for bound states and thus the fuzzy sphere with
angular momentum is still unstable.
In any case, we would not anticipate the configuration to be stable as the D-branes can emit their Ramond-Ramond
charge via synchrotron emission \cite{radiation}. It would be useful to find an ansatz to allow for the inclusion of $p=6$ backgrounds
as there is the possibility of a bound state in that case, however this may require uplifting to M-theory.

The non-BPS system in both $Dp$-brane and $NS$5-brane background leads to richer dynamics than in the BPS case, thanks to the
existence of world-volume symmetries. In both cases we looked at the classical solutions when the tachyon field 
affects the strength of the gravitational attraction of the branes to the
background. From the Abelian field theory description of unstable branes, we know that as $T \to \pm \infty$ the open string
modes are confined leading to the destruction of the brane and the appearance of a stringy fluid \cite{tachyon_stuff}. The dual picture should give
us some insight into what happens in the non-Abelian theory, and whether the Open/Closed duality conjecture remains valid.
 We used the symmetry of the fields to explicitly examine the $D3$-backgrounds, but it
would be useful to extend the work of Kluson \cite{non_bps_dynamics} to the full symmetry transformation for general $Dp$-brane
backgrounds. This additional symmetry has profound effects on the dynamics of the fuzzy sphere in the near horizon geometry,
however we do not know if the symmetry even exists in the quantum theory. 

We have briefly examined the dual version of the QHS and found agreement \cite{susskind, hyakutake} with aspects of the Abelian theory,
namely that the stabilisation radius of the system is almost identical. The origin of the string contribution is 
unclear in  our non-Abelian construction
and so we have only constructed the dual picture to the dielectric effect, namely the Myers effect. Furthermore, we have shown that
the equations of motion in the two pictures are identical in a curved background  when we take the large $N$ limit of the symmetrized trace.
It would be useful to extend the work initiated here to the study of the non-Abelian QHS model and compare the results to those
obtained using matrix theory. Finally we have investigated higher (even) dimensional fuzzy spheres in the $Dp$-brane background and found
that only collapsing solutions are admissible. The case of odd dimensional fuzzy spheres is non-trivial and certainly merits future investigation.
In addition, it would be interesting to try and construct the dual pictures of these collapsing solutions to see if the equations of motion are
identical in the Abelian theory. This is complicated by many factors, for example the classical limit of the fuzzy $4$-sphere is six dimensional 
because the algebra belongs to the coset $SO(5)/U(2)$. We must project out the $U(2)$ states in order to construct the dual picture.
We hope that this note has provided some small insight into
the dynamics of fuzzy spheres in selected curved backgrounds, and we hope to return to some of these problems in the future.

\begin{center}
\textbf{Acknowledgements.}
\end{center}
Thanks go to C. Papageorgakis, S. McNamara, J. Bedford and S. Ramgoolam for their useful insights and comments.
JW is supported by a QMUL studentship, and thanks the theoretical physics group at
Stockholm University for its kind hospitality.
This work was in part supported by the EC Marie Curie Research Training Network
MRTN-CT-2004-512194.

\end{document}